\documentclass[journal,twoside,web]{ieeecolor}
\usepackage{generic}
\usepackage{cite}
\usepackage{amsmath,amssymb,amsfonts}
\usepackage{algorithmic}
\usepackage{graphicx}
\usepackage{algorithm,algorithmic}
\usepackage{hyperref}
\hypersetup{hidelinks=true}
\usepackage{textcomp}

\usepackage{mathrsfs}
\usepackage{booktabs}
\usepackage{subcaption}

\def\BibTeX{{\rm B\kern-.05em{\sc i\kern-.025em b}\kern-.08em
  T\kern-.1667em\lower.7ex\hbox{E}\kern-.125emX}}
\begin{document}
\title{An Autocovariance Least-Squares-Based Data-Driven Kalman Filter for Unknown Systems}
\author{Suyang Hu, Xiaoxu Lyu, Peihu Duan, Dawei Shi, and Ling Shi
\thanks{Suyang Hu, Xiaoxu Lyu, and Ling Shi are with the Department of Electronic and Computer Engineering, Hong Kong University of Science and Technology, Hong Kong, China (e-mail: shuat@connect.ust.hk; eelyuxiaoxu@ust.hk; eesling@ust.hk). }
\thanks{Peihu Duan is with the School of Electrical Engineering and Computer Science, KTH Royal Institute of Technology, Stockholm, Sweden (e-mail: peihu@kth.se).}
\thanks{Dawei Shi is with the State Key Laboratory of Intelligent Control and Decision of Complex Systems, School of Automation, Beijing Institute of Technology, Beijing 100081, China (e-mail: daweishi@bit.edu.cn).}
 }

\maketitle

\begin{abstract}
This article investigates the problem of data-driven state estimation for linear systems with both unknown system dynamics and noise covariances. We propose an Autocovariance Least-squares-based Data-driven Kalman Filter (ADKF), which provides a unified framework for simultaneous system identification and state estimation by utilizing pre-collected input-output trajectories and estimated initial states. Specifically, we design a SDP-based algorithm for estimating the noise covariances. We quantify the impact of model inaccuracy on noise covariances estimation using this identification algorithm, and introduce a feedback control mechanism for data collection to enhance the accuracy and stability of noise covariance estimation. The estimated noise covariances account for model inaccuracy, which are shown to be more suitable for state estimation. We also quantify the performance gap between the ADKF and the traditional Kalman filter with known system dynamics and noise covariances, showing that this gap decreases as the number and length of pre-collected trajectories increase. Finally, numerical simulations validate the robustness and effectiveness of the proposed ADKF. 
\end{abstract}


\section{Introduction}
\label{sec:introduction}
\IEEEPARstart{A}{ccurate} state estimation is essential for effective control, monitoring, and advanced information extraction from real-world systems \cite{simon2006optimal}. The Kalman filter \cite{kalman1960new}, a foundational tool for state estimation in noisy environments, has been widely applied in fields such as aerospace \cite{kim2007kalman}, control systems \cite{10313083}, and computer vision \cite{zhou2020kfnet}. Its optimal performance in minimizing the mean squared estimation error depends heavily on accurate prior system models, and model mismatches can significantly degrade the performance of the filter~\cite{tsiamis2022online}. However, obtaining precise prior models based on first principles can be challenging in many practical scenarios~\cite{bhagwat2021computation}~\cite{xue2020system}. To address this challenge, recent research has increasingly resorted to data-driven filtering methods \cite{hou2013model}, which utilize pre-collected data and online data to estimate the state of unknown systems directly. A critical issue for the data-driven filtering approach is how to learn a reliable filter from noisy data while maintaining the feasibility and optimality. This article seeks to address this challenge.

\subsection{Related Works}
According to the design criterion, data-driven filters reported in the existing literature are generally categorized into two main approaches: indirect and direct data-driven filters. 
\vspace{6pt}

Indirect data-driven filters involve a two-step process. First, a partial or complete system model is identified from the collected data using system identification techniques, such as optimization algorithms \cite{lai2021capacity}\cite{duan2023data}, and neural networks \cite{coskun2017long} \cite{rangapuram2018deep}. Once the model is identified, a filter is designed based on it. For example, Xin et al.~\cite{lai2021capacity} integrated a particle swarm optimization algorithm with an extended Kalman filter to improve the parameter estimation of a battery's Arrhenius aging model, leading to more accurate capacity estimation for lithium-ion cells. Similarly, Duan et al.~\cite{duan2023data} proposed a data-driven Kalman filter for unknown linear systems by constructing a state-space model from noisy data, while also ensuring the filtering performance. Neural network-based methods have also been explored; for example, Syama et al.~\cite{rangapuram2018deep} employed a recurrent neural network to parametrize a linear state-space model for robust time series forecasting. The identification of the system model not only improves our understanding of system characteristics but also enables more flexible filter design. This is particularly significant in fields such as aerospace \cite{liu2019fuel} and robotics \cite{qin2018vins}, where interpretability is crucial.
\vspace{6pt}

In contrast, direct data-driven filters bypass model identification and operate solely on the available data, using techniques such as the Set-Membership framework~\cite{milanese2010direct}~\cite{milanese2009filter}, adaptive algorithms \cite{shafieezadeh2018wasserstein}, Willems’ fundamental lemma~\cite{willems2005note,yin2021maximum, turan2021data, liu2024learning}, and weighted least-squares optimization \cite{zamzam2019data}. 
While these methods can achieve strong performance under certain prior assumptions, they may encounter challenges such as overfitting and sensitivity to noise, which may restrict their practical applicability. For example, Milanese et al. \cite{milanese2010direct} proposed a direct data-driven filter within a Set Membership framework, which has been shown to be almost-optimal under bounded process noise. Liu et al. \cite{liu2024learning} developed an SDP-based algorithm to directly compute the steady-state Kalman gain based on Willems' fundamental lemma; however, it requires bounded noise to ensure the stability and feasibility of the method. Additionally, weighted least-squares-based methods have been applied to distribution system state estimation in a data-driven framework. For example, Ahmed et al. \cite{zamzam2019data} combined a shallow neural network with the weighted least-squares method to initialize Gauss-Newton algorithms for distribution system state estimation, improving both convergence speed and reliability. However, weighted least-squares-based methods are more suitable for constant parameters than states of dynamic systems.
 
\subsection{Motivations}
Most indirect data-driven filters primarily focus on recovering unknown system dynamics, while assuming that noise covariances are known. Additionally, in practical applications, noise covariances are also typically tuned empirically without accurate estimation \cite{lai2021capacity}\cite{duan2023data}. These scenarios can lead to inaccuracy in the use of noise covariances. However, noise covariances are crucial for accurate state estimation, and inaccuracies in their values can significantly degrade filtering performance \cite{mehra1972approaches}. Furthermore, manually tuning noise covariances is time-consuming and difficult, often leading to suboptimal state estimation performance. Therefore, improving the robustness and reliability of data-driven filters requires addressing both unknown system dynamics and noise covariances.
\vspace{6pt}

Various methods have been developed for estimating noise covariances, including correlation methods \cite{ rajamani2009estimation}, covariance matching~\cite{solonen2014estimating}, maximum likelihood \cite{schon2011system}, and Bayesian methods \cite{huang2017novel}. A detailed overview of these methods, including their properties, advantages, and limitations, can be found in \cite{mehra1970identification}\cite{odelson2006new}. However, most of these techniques assume known system dynamics and then estimate the noise covariances based on those dynamics. For example, the autocovariance least-squares (ALS) method \cite{rajamani2009estimation}\cite{odelson2006new}, a widely used correlation-based technique that forms the basis of our approach, assumes known system dynamics, and can provide unbiased estimation of noise covariances based on those dynamics. However, applying ALS to unstable systems with inaccurate dynamics can result in failed noise covariance estimation. Consequently, several new challenges arise when designing data-driven filters, particularly in scenarios where both system dynamics and noise covariances must be identified simultaneously:
\begin{enumerate}
	\item  The inherent coupling between system dynamics and noise covariances makes estimation of the latter challenging when system dynamics are unknown. 
		\vspace{6pt}
		
	\item Designing a data-driven Kalman filter that accounts for both unknown system dynamics and noise covariances is complex and has not been extensively explored.
	\vspace{6pt}
	
	\item Quantitatively analyzing the coupled effects of system dynamic inaccuracies on noise covariance estimation and filtering performance is a further unresolved challenge. 
\end{enumerate}

\subsection{Contributions}
 Inspired by the above discussion, this paper investigates the state estimation problem for linear systems with both unknown system dynamics and noise covariances. We integrate system identification with state estimation, and propose an integrated ALS-based data-driven Kalman filter (ADKF). Additionally, we derive a sample-complexity bound that accounts for the combined effects of inaccuracies in both the system dynamics and noise covariances on the performance of the proposed ADKF. Compared to the existing literature, this paper has three key advantages:
\begin{enumerate}
	\item We propose a data-driven framework that jointly estimates system dynamics and noise covariance matrices for unknown linear systems, focusing on the impact of system dynamics inaccuracies on noise covariances estimation. A feedback mechanism is integrated into the data collection process to ensure both the stability and accuracy of the ALS method (\textbf{Proposition~1}). 
	\vspace{6pt}

	\item We analyze the relationships among the nominal, true, and estimated noise covariance matrices, which allows for an evaluation of the performance of the proposed ADKF under different nominal noise covariance matrices. Additionally, we demonstrate how the estimation accuracy for noise covariance matrices depends on the sampling sequence length (\textbf{Theorem~1}, \textbf{Theorem~2}).
	\vspace{6pt}
	
	\item We quantitatively analyze the filter’s performance and establish its relationship with the number of samples and the sampling sequence length (\textbf{Theorem~3}). This analysis reveals that as the number of samples and sequence length increase, the performance of the data-driven estimator approaches that of the optimal filtering.
\end{enumerate}
\vspace{6pt}

The remainder of this article is organized as follows. Section~\ref{sec3} presents the problem formulation. Section~\ref{sec4} explains the principles of the ADKF. Section~\ref{sec5} analyzes the filtering performance of the ADKF. Section~\ref{sec6} provides numerical examples to validate the properties of the ADKF. Section~\ref{sec7} concludes the article.
\vspace{6pt}

\textit{Notations:} Let $\mathbb{N}$ denote the set of natural numbers. Let $\mathbb{R}^n$ and $ \mathbb{R}^{n \times m}$ denote the sets of $n$-dimensional column vectors and $(n*m)$-dimensional real matrices, respectively. Let $0$ denote the scalar, vector, or matrix of the corresponding dimension, with all elements equal to zero. Let $I_n$ denote the $n$-order identity matrix. Let $A \succ 0$ ( $ A \succeq 0$) denote the positive definite (semidefinite) matrix. Let the operator $||\cdot||_2$ denote the 2-norm for matrices and the Euclidean norm for
vectors. Let the operator $||\cdot||_{F}$ denote the Frobenius norm for matrices. Let $\otimes$ denote the Kronecker product. Let $\mathcal{N}(\mu, \Sigma)$ denote the Gaussian distribution with mean $\mu$ and covariance $\Sigma$. Let $A^{\dagger}$ denote the Moore–Penrose inverse of $A$. Given $x \in \mathbb{R}^n$, let $\text{cov}(x)$ denote the covariance of the vector $x$, where $\text{cov}(x) = E\{ (x - E\{x\}) (x - E\{x\})^T\}$. Let $\text{vec}(A) \in \mathbb{R}^{n^2}$ denote the vectorization of $A \in \mathbb{R}^{n \times n}$, and $\text{vech}(A) \in \mathbb{R}^{\frac{n(n+1)}{2}}$ denote the half-vectorization, which stacks the column-wise elements of the lower triangular part of $A$. If $\mathbf{A}$ is a block matrix with diagonal blocks denoted as $A_{ii}$, then define $\text{tr}_{\text{block}}(\mathbf{A}) = \sum A_{ii}$. Given two integers $a,b \in \mathbb{N}$ with $a<b$, let $\boldsymbol{x}^i_{[a,b]}$ denote the $i$-th data, where $\boldsymbol{x}^i_{[a,b]} \triangleq [ x_{i,a}^T, \ldots,  x_{i,b}^T ]^T$.

\section{Problem Formulation \label{sec3}}

\subsection{System Model}
Consider a discrete-time linear time-invariant (LTI) system 
\begin{equation}
	\begin{aligned}
		x_{k+1} &= Ax_k + Bu_k + \omega_k, \\
		y_k & = Cx_k + \nu_k, \quad k \in \mathbb{N},
		\label{eq:LTImodel}
	\end{aligned}
\end{equation}
where $x_k \in \mathbb{R}^n$, $u_k \in \mathbb{R}^m$, and $y_k \in \mathbb{R}^p$ represent the system state, control input, and measurement output at time step $k$, respectively. The process noise $\omega_k  \in \mathbb{R}^n \sim \mathcal{N}(0, Q)$ and measurement noise $\nu_k \in \mathbb{R}^p \sim \mathcal{N}(0, R)$ are mutually uncorrelated, zero-mean white Gaussian, with $Q \succeq 0$ and $R \succ 0$. 
\vspace{6pt}

\noindent \textbf{Assumption 1.} The pair $(A,C)$ is observable, and $(A, \sqrt{Q})$ is controllable. 
\vspace{6pt}

The above assumption is primarily made to ensure the convergence of the Kalman filter \cite{kalman1961new}, which is a standard requirement. We assume that the system matrices $A,B,C$, and the noise covariance matrices $Q$ and $R$ are unknown. Since different choices of state variables can result in distinct state-space models, we further assume that the system states are fixed and organized into a vector corresponding to a specific state-space model. 

\subsection{Data Collection \label{data_collection}}
Suppose that we conduct $N$ independent experiments. In the $i$-th experiment, the initial state is set as $\hat{x}_{i,0}$, which is affected by an uncorrelated and unknown Gaussian noise $\Delta {x}_{i,0} \sim \mathcal{N}(0, \Pi_0) $. Therefore, we can regard $\hat{x}_{i,0}$ as an estimation of the true initial state $x_{i,0}$, expressed as $x_{i,0} = \hat{x}_{i,0} + \Delta {x}_{i,0} $. We then set the input sequence $\boldsymbol{u}^i_{[0,\tau - 1]}$ and collect the corresponding output sequence $\boldsymbol{y}^i_{[0,\tau]}$, where $\tau$ represents the sequence length. Importantly, both $\hat{x}_{i,0} $ and $u_{i, k} $ in $\boldsymbol{u}^i_{[0,\tau - 1]}$ are modeled as Gaussian distributions, specifically $ \hat{x}_{i,0} \sim \mathcal{N}(0, \sigma_x^2 I_n)$ and $u_{i, k} \sim \mathcal{N}(0, \sigma_u^2 I_m)$ for $k=0, \dots, \tau -1 $, where $\sigma_x$ and $\sigma_u$ are user-defined constants. These assumptions are essential for analyzing filtering performance but do not affect the filter design, as discussed in Section~\ref{sec5}. Consequently, the available data from all $N$ trajectories can be combined into data matrices as follows
\begin{subequations} \label{eq:data_matrix}
	\begin{align}
	    &\hat{X}_0 =\left[ \hat{x}_{1,0} ,\ldots , \hat{x}_{N,0} \right] ,\label{eq:x_hat_0}\\
		&Y =\left[ \boldsymbol{y}^1_{[0,\tau ]} , \ldots , \boldsymbol{y}^N_{[0,\tau ]} \right],\label{eq:YYY} \\
		&U = \left[ \boldsymbol{u}^1_{[0,\tau - 1]}, \ldots, \boldsymbol{u}^N_{[0,\tau - 1]} \right].\label{eq:UUU} 
	\end{align}
\end{subequations}

To estimate the noise covariance matrices, we conduct additional $N_0$ experiments, initializing the state based on the previously defined distribution. A key difference in these experiments is the introduction of an output feedback controller, where $u_k = K_c y_k, k=0,\ldots, \tau-1$. The nominal data-driven Kalman filter (NDKF), as proposed in \cite{duan2023data}, is then applied using the estimated system matrices $\hat{A}$ and $\hat{{B}}$, along with nominal mismatched noise covariance matrices $Q_u$ and $R_u$. We collect the innovation sequence $\boldsymbol{z}^i = [z^i_1,\ldots,z^i_{\tau}]$ for $i=1,\ldots,N_0$, where $z^i_k = y^i_{k} - C\hat{x}^i_{k|k-1}, k=1,\ldots, \tau$, and $\hat{x}^i_{k|k-1}$ denotes the prior estimate of $x_k^{i}$. The corresponding error covariance of $\hat{x}^i_{k|k-1}$ is represented as $P^i_{k|k-1} = E\{ (x^i_k - \hat{x}^i_{k|k-1} ) (x_k - \hat{x}^i_{k|k-1} )^T \}$. Thus, an additional data matrix $Z$ is formed for estimating the noise covariance matrices as follows 
\begin{equation}
		\begin{aligned}
			Z =  \begin{bmatrix}
				({\boldsymbol{z}}^1)^T, \dots, ({\boldsymbol{z}}^{N_0})^T \end{bmatrix}^T. \label{eq:collect_z}
		\end{aligned}
\end{equation} 

The feedback controller is essential for estimating the noise covariance matrices because, if system matrix $A$ is not Schur stable, the noise covariance matrices to be estimated will diverge. Conversely, if $A$ is stable, the existing data \eqref{eq:data_matrix} can be utilized for noise covariance estimation, thus eliminating the need for extra experiments. Later in Section~\ref{gsencm}, we will discuss on this result. Moreover, it is worth mentioning that the values of $Q_u$ and $R_u$ do not affect the estimation of the noise covariance matrices, as analyzed in Theorem~1. 
\vspace{6pt}

Finally, denote the corresponding unknown initial state, process, and measurement noise data as follows
\begin{equation}
	\begin{aligned}
		&\tilde {X}_0 = \left[
		\Delta x_{1,0} , \dots , \Delta x_{N,0}
		\right],\\
		&V = \left[\boldsymbol{\nu}^1_{[0,\tau]} , \dots ,\boldsymbol{\nu}^N_{[0,\tau]}  \right] , \\ 
		&\Omega = \left[
		\boldsymbol{\omega}^1 _{[0,\tau-1]} , \dots , \boldsymbol{\omega}^N_{[0,\tau-1]} 
		\right]. \\
		\label{eq:unknown_noise}
	\end{aligned}
\end{equation}
Based on \eqref{eq:LTImodel}, the data matrices in \eqref{eq:data_matrix} and \eqref{eq:unknown_noise} can be written as follows
\begin{equation*}
	Y = G \hat{X}_0 + H(I_{\tau} \otimes B) U + G \tilde {X}_0 + H\Omega + V,
	\label{eq:Matrix_form of Y}
\end{equation*}
where
\begin{equation}
	G =\begin{bmatrix}
		C\\
		CA\\
		\vdots\\
		CA^{\tau} 
	\end{bmatrix}, \quad H = \begin{bmatrix}
		0 & 0 & \dots & 0 \\
		C & 0 & \dots & 0 \\
		CA & C & \dots & 0 \\
		\vdots & \vdots & \ddots & \vdots \\
		CA^{\tau-1} & CA^{\tau-2} & \dots & C
	\end{bmatrix}.
\end{equation}

\noindent \textbf{Assumption 2.} $\tau \geq n$.\\

\noindent \textbf{Assumption 3.} $\text{rank} \begin{bmatrix}
	\hat{X}_0\\
	U
\end{bmatrix}= n + \tau m$. \\

\noindent \textbf{Remark 1.} Assumption 2 is essential for ensuring that the observability matrix included in $G$ has full column rank. Assumption~3 establishes a persistent excitation condition, similar to that in system identification and Willems’ fundamental lemma \cite{duan2023data}. Given that $\hat{X}_0$ and $U$ are both generated by a random process, Assumption~3 can be easily satisfied. If this is not the case, we can add small Gaussian noise to $\hat{x}_{i,0}$. This is acceptable because $\hat{x}_{i,0}$ serves as an estimation of $x_{i,0}$, and adding a small value will not significantly impact the results. Additionally, Assumption~3 can also be satisfied when the number of trajectories meets the criterion $ N \gg n + \tau m$, with larger $N$ leading to improved filtering performance, as will be discussed in Section~\ref{sec5}.

\subsection{Problem of Interest}
The objective of this article is to design a filter algorithm for estimating the state of linear systems \eqref{eq:LTImodel} with unknown system matrices $A,B,C$, and mismatched noise covariances $Q_u, R_u$ by utilizing pre-collected data \eqref{eq:data_matrix}, \eqref{eq:collect_z}, and online data, denoted as $\{ \boldsymbol{u}^{o}_{[0,k]}, \boldsymbol{y}^{o}_{[0,k]}\}$. The following specific objectives are addressed:
\vspace{6pt}
\begin{enumerate}
	\item Reveal the effects of system dynamic inaccuracies and mismatched noise covariances $Q_u, R_u$ on state estimation. Design a stable and precise system identification method $h(\cdot)$ for jointly identifying the system model and noise covariance matrices as
	\begin{equation*}
		(\hat{A}, \hat{B},\hat{C},\hat{Q},\hat{R} ) = h(\hat{X}_0, U, Y, Z, Q_u, R_u).
	\end{equation*}
		
	\item Design a data-driven filter $g(\cdot)$ to estimate the state $x_k$ of unknown linear systems based on the identified system model. The filter can be mathematically formulated as
	\begin{equation*}
		\begin{aligned}
				\hat{x}_{k|k} = g( \boldsymbol{u}^o_{[0,k-1]}, \boldsymbol{y}^o_{[0,k]}, \hat{A}, \hat{B},\hat{C},\hat{Q},\hat{R} ).
			\end{aligned}
	\end{equation*}
	where $\hat{x}_{k|k}$ represents the posterior estimate of the state at time step $k$.  
\vspace{6pt}

	\item Conduct a quantitative analysis and comparison of the performance of the proposed ADKF algorithm with that of the traditional Kalman filter with accurate system models.
\end{enumerate}
\vspace{6pt}

\section{ALS-Based Data-driven Kalman Filter \label{sec4}}
In this section, we outline the principles of the proposed ADKF algorithm. We begin by detailing the procedure for identifying the system matrices ${{A}}$, ${B}$, and $C$ using the pre-collected data \eqref{eq:data_matrix}. Next, we discuss the conditions that ensure stable estimation of the noise covariance matrices. Finally, we present the overall framework of the ADKF algorithm.

\subsection{System Model Identification \label{sid}}
We first estimate the system matrices $A,B$, and $C$ from the pre-collected data \eqref{eq:data_matrix}, denoted as $\hat{A}$, $\hat{B}$, and $\hat{C}$, by
\begin{equation}
	\begin{aligned}
			&\hat{A} = 	{G}_1^{\dagger} {G}_2, \ \hat{B} = 	{G}_1^{\dagger} {G}_3, \ \hat{C} = 	G_1(1:p; 1:n),
	\end{aligned}
	\label{eq:est_model}
\end{equation}
where
\begin{equation}
	\begin{aligned}
		&{G}_1 = \left( Y \begin{bmatrix}
			\hat{X}_0\\
			U
		\end{bmatrix}^{\dagger} \right) (1:\tau p; 1: n) ,\\
		&{G}_2 = \left( Y \begin{bmatrix}
			\hat{X}_0\\
			U
		\end{bmatrix}^{\dagger} \right) (p +1:(\tau+1) p; 1: n) ,\\
		&{G}_3 = \left( Y \begin{bmatrix}
			\hat{X}_0\\
			U
		\end{bmatrix}^{\dagger} \right) (p+1:(\tau+1) p; n+1: n+m) .\\
	\end{aligned}
	\label{eq:g1g2g3}
\end{equation}

A unique estimation of the system matrices $\hat{{A}}$, $\hat{B}$ and $\hat{{C}}$ requires the uniqueness of $Y ([\hat{X}_0^T \ \ U^T]^T)^{\dagger}$, which is easily satisfied under Assumption 3. Additionally, the matrix $G_1$ must have full column rank. As shown in \cite[Proposition 2]{duan2023data}, there exists a constant $N_1$ such that, for $N > N_1$, $G_1$ has full column rank with probability $1-\delta$, where $\delta \in (0,1)$ is any positive scalar. The value of $N_1$ depends on $\delta$ and system noise. Consequently, $G_1$ will have full column rank with high probability if the conditions mentioned earlier are satisfied. Thus, $N_1$ provides a benchmark for determining the necessary number of pre-collected trajectories. 

\subsection{Guarantees for Stable Estimation of Noise Covariance Matrices \label{gsencm}  }
To accurately estimate the system state, the unknown noise covariance matrices must be determined. This is crucial for two reasons: first, the accuracy of these matrices directly affects filtering performance; second, mismatched noise covariances can lead to underestimation of $N_1$, resulting in failure to satisfy the full-rank condition of $G_1$, thereby producing an inaccurate system model.
 \vspace{6pt} 
 
In this article, we apply the ALS method to estimate the noise covariance matrices. However, directly applying the ALS method with the estimated matrices $\hat{A}$, $\hat{B}$, and $C$ will fail if the original system matrix $A$ is unstable. To analyze this potential issue, let $\Delta A= \hat{{A}} - A$, $\Delta B= \hat{{B}} - B$, and $\Delta C= \hat{{C}} - C$. The system \eqref{eq:LTImodel} can then be rewritten as follows
\begin{equation}
	\begin{aligned}
		x_{k+1} & = \hat{A} x_k + \hat{B} u_k + \hat{\omega}_k,\\
		 y_k & = \hat{C} x_k + \hat{\nu_k}, \quad k \in \mathbb{N},
	\end{aligned}
\end{equation}
where $\hat{\omega}_k = \Delta A x_k + \Delta B u_k +\omega_k$ and $\hat{\nu_k} = \Delta C x_k + \nu_k$. Since the initial state $x_0$ and input $u_k$ in \eqref{eq:data_matrix} follow zero-mean Gaussian distributions, both $\hat{\omega}_k$ and $\hat{\nu_k} $ can be regarded as linear combinations of zero-mean Gaussian distributed variables. Therefore, the covariance matrix $\tilde{Q} $ of $\hat{\omega}_k$ is defined as follows
\begin{equation}
	\begin{aligned}
		\tilde{Q} & = E\{\hat{\omega}_k\hat{\omega}_k^T\} = \Delta \tilde{Q}_k + Q,
	\end{aligned}
	\label{eq:QTK}
\end{equation} 
where
\begin{equation}
	\begin{aligned}
	&	\Delta \tilde{Q}_k
		= (\Delta A) E\{x_{k} x_{k}^T\} (\Delta A)^T + (\Delta B) E\{u_{k} u_{k}^T\} (\Delta B)^T.
	\end{aligned}
	\label{eq:delta_q_k}
\end{equation}
For the term $E\{x_{k} x_{k}^T\}$, based on \eqref{eq:LTImodel} and \eqref{eq:data_matrix}, we have
\begin{equation}
	\begin{aligned}
		E\{x_{k} x_{k}^T\} 
		& = \sum_{i=1}^{k} A^{i-1}  (\sigma_u B B^T + Q) (A^{i-1})^T \\
		&\quad +  A^k (\Pi_0 + \sigma_x I_n) (A^k)^T .
	\end{aligned}
	\label{eq:exkxk_no}
\end{equation}

As shown in \eqref{eq:delta_q_k} and \eqref{eq:exkxk_no}, the term $ \Delta \tilde{Q}_k $ is significantly affected by the accuracy of the system model and the stability of $A$. Due to inevitable deviations between the estimated and true system models, $ \Delta \tilde{Q}_k $ cannot be eliminated. If $A$ is unstable, the term $A^k$ in \eqref{eq:exkxk_no} will diverge to infinity as $k \to \infty$, causing $ \Delta \tilde{Q}_k \to \infty $. Furthermore, as $k \to \infty$, the series summation term in $E\{ x_kx_k^T\}$ will also diverge. These two effects can lead to the failure of estimating $\tilde{Q}$. This instability also arises in the covariance matrix of $\hat{\nu_k}$, denoted as $\tilde{R} = E \{\hat{\nu_k}\hat{\nu_k}^T \}$.
 \vspace{6pt} 
 
To mitigate this instability, we design a feedback controller $u_k = K_c y_k$ that ensures ${A} + {B} K_c C$ is Schur stable. Subsequently, data \eqref{eq:collect_z} is collected from additional $N_0$ trajectories as described in Section \ref{data_collection}. Under the stable controller gain $K_c$, we obtain the following important result about $\tilde{Q}$ and $\tilde{R}$.
 \vspace{6pt} 
 
 \noindent \textbf{Proposition 1.} Under Assumptions 1 - 3, if there exists a stable output feedback controller gain $K_c$, such that ${A} + {B} K_c C$ is Schur stable, then as $k \to \infty$, we have
 \begin{equation}
 	\begin{aligned}
 		||\tilde{Q} - Q||_2 & = || \Pi_2 ||_2 \le l_1 ||Q||_2, \\
 		|| \tilde{R} - R||_2 & =  || \Pi_3 ||_2 \le l_2 ||Q||_2, \\
 	\end{aligned}
 \end{equation}
where $l_1,l_2$ are constants, and $\Pi_2, \Pi_3$ are constant matrices defined as
\begin{equation}
	\begin{aligned}
		&\Pi_1
		=  \sum_{i=1}^{\infty} (A+BK_c C)^{i-1} Q [(A + BK_c C)^{i-1 }]^T ,\\
		&\Pi_2 = (\Delta A + \Delta B K_c C) \Pi_1 (\Delta A + \Delta B K_c C)^T, \\
		&\Pi_3 = 	(\Delta C) \Pi_1 ( \Delta C)^T .
	\end{aligned}
	\label{eq:p123}
\end{equation}

\begin{proof}
 See the proof in Appendix \ref{pppp1}.\\
\end{proof}

 Some existing works \cite{dean2020sample}\cite{ de2021low} provide methods for designing the controller gain $K_c$ using pre-collected data. In this article, $K_c$ can be computed via the equation $K_c \hat{C} M_1 = M_2$, where $M_1$ and $M_2$ are user-defined constant matrices. The gain $K_c$ designed in this way is proven to stabilize the system \cite[Theorem 2]{duan2023data}. Furthermore, under Proposition~1, when $k \to \infty$, the noise $\hat{\omega}_k$ and $\hat{\nu}_k$ follow Gaussian distributions, such that $\hat{\omega}_k \sim \mathcal{N}(0, \tilde{Q})$ and $\hat{\nu}_k \sim \mathcal{N}(0, \tilde{R})$. Here, $\tilde{Q}$ and $\tilde{R}$ are unknown constant matrices that need to be estimated. 
  \vspace{6pt} 
  
 \noindent \textbf{Remark 2.} As shown in \eqref{eq:p123}, the controller gain $K_c$ can affect the value of $\Pi_1$, thereby affecting the values of $\tilde{Q}$ and $\tilde{R}$. Therefore, to obtain $\tilde{Q}$ and $\tilde{R}$ that closely approximate the true $Q$ and $R$, $K_c$ should be designed to be as small as possible. We believe that the joint design of state estimation and LQG control can identify the optimal controller gain $K_c^*$. A detailed investigation is left for future work.
 
 \subsection{The ALS-Based Data-Driven Kalman Filter\label{als_fram} }
In this subsection, we will outline the principles of ADKF algorithm. First, we estimate the noise covariance matrices based on the estimated system model obtained from \eqref{eq:est_model}. 
 \vspace{6pt} 
 
The prediction error $\varepsilon_k = x_k - \hat{x}_{k|k-1}$ evolves as follows
\begin{equation}
	\varepsilon_{k+1} = \underbrace{(\hat{A} - \hat{A}K_f\hat{C})}_{\bar{A}} \varepsilon_k + \underbrace{[I_n, -\hat{A} K_f]}_{\bar{\Gamma}} \underbrace{\begin{bmatrix} \hat{w}_k \\ \hat{v}_k \end{bmatrix}}_{\bar{w}_k},
	\label{eq:simplified_ek1}
\end{equation}
where $K_f$ is the stable Kalman gain from the NDKF. Let $\bar{Q} = E \{\bar{\omega}_k \bar{\omega}_k^T\}$, and express the autocovariance as follows
\begin{equation}
	\begin{aligned}
E\{z_{k+j} z_k^T\} = \begin{cases}
	\hat{C} \bar{P}_f \hat{C} ^T + \tilde{R},& j=0,\\
	\hat{C} \bar{A}^j \bar{P}_f \hat{C} ^T - \hat{C} \bar{A}^{j-1}	\hat{A} K_f \tilde{R}, &j \ge 1 ,
\end{cases}
	\end{aligned}	
	\label{eq:autocovariance}
\end{equation}
where $z_k = y_k - C\hat{x}_{k|k-1}$, and $\bar{P}_f = E\{\varepsilon_{k} \varepsilon_{k}^T \}$ is the stable prior estimation error covariance, obtained by solving the Lyapunov equation
\begin{equation}
	\bar{P}_f = \bar{A} \bar{P}_f \bar{A} ^T + \bar{\Gamma} \bar{Q} \bar{\Gamma}^T.
\end{equation}
 
Let $\hbar_j = E\{z_{k+j} z_k^T\}$ for $ j=0,\dots,L-1$, where $L$ is a user-defined hyperparameter. We can then form $ \Upsilon_{L} $ as follows
\begin{equation}
	\begin{aligned}
		\Upsilon_{L} &= \begin{bmatrix}
			\hbar_0^T, & \hbar_1^T, & \hbar_2^T , & \dots , & \hbar_{L - 1}^T \\
		\end{bmatrix}^T \\
		&= \mathcal{G}_o \bar{P}_f \hat{C}^T + \mathcal{H} \tilde{R},
	\end{aligned}
	\label{eq: autocovariance_matrix}
\end{equation}
where 
\begin{equation}
	\begin{aligned}
		&\mathcal{G}_o = \begin{bmatrix}
			\hat{C} \\
			\hat{C} \bar{A} \\
			\vdots \\
			\hat{C} \bar{A}^{L - 1}
		\end{bmatrix}, \quad
		\mathcal{H} = \begin{bmatrix}
			I_p \\
			-\hat{C} \hat{A} K_f \\
			\vdots \\
			-\hat{C} \bar{A}^{L - 2} \hat{A}K_f
		\end{bmatrix} .
	\end{aligned}
\end{equation}
Applying the vectorization operator to $\Upsilon_{L}$, we obtain
\begin{equation}
	\begin{aligned}
		d &= \text{vec} (\Upsilon_{L}) \\
		& = [ (\hat{C} \otimes \mathcal{G}_o )(I_{n^2} - \bar{A} \otimes \bar{A})^{-1} ] \text{vec}(\tilde{Q}) \\
		&\quad + [ (\hat{C} \otimes \mathcal{G}_o )(I_{n^2} - \bar{A} \otimes \bar{A})^{-1} (\hat{A}K_f \otimes \hat{A}K_f )\\
		&\quad  + I_{p} \otimes \mathcal{H} ] \text{vec} (\tilde{R}).
	\end{aligned}
	\label{eq:vec_autocovariance}
\end{equation}

Next, we will utilize the data \eqref{eq:collect_z} to estimate $	\text{vec} (\Upsilon_{L}) $. Before proceeding, we introduce the following assumption.
\vspace{6pt}

\noindent \textbf{Assumption 4.} The $K_f$-innovations data $z_k, k=1,\dots,\tau_1$, are obtained after the filter has reached steady state.
 \vspace{6pt} 
 
To fulfill Assumption 4, we extract the last $\tau_1$ columns of the data matrix \eqref{eq:collect_z}, denoted as $Z_s $, where
\begin{equation}
	Z_s = \begin{bmatrix}
		{z}_{b+1}^1 & {z}_{b+2}^1 & \dots & {z}^1_{b+ \tau_1 } \\
		{z}_{b+ 1}^2 & {z}_{b+ 2}^2 & \dots & {z}^2_{b+ \tau_1 } \\
		\vdots & \vdots & \ddots & \vdots\\
		{z}_{b+ 1}^{N_0} & {z}_{b+ 2}^{N_0} & \dots & {z}^{N_0}_{b+ \tau_1} 
	\end{bmatrix}, \label{eq:newZ}
\end{equation}
with $b = \tau - \tau_1$. Since \eqref{eq:autocovariance} is independent of $k$, we utilize $Z_s$ to estimate the noise covariance matrices. For $j = 0,1,\cdots,L-1$, define
\begin{equation}
	\begin{aligned}
		\mathcal{Z}_j &= Z_s(1:p{N_0}, j+1:\tau_1)Z_s(1:p{N_0}, 1:\tau_1 - j)^T\\
		&=\begin{bmatrix}
			\sum_{q=b+1}^{b+\tau_1-j} z^1_{q+j} (z^1_{q})^T &  \dots & \dots \\
			\vdots & \ddots & \vdots\\
			\dots & \dots & \sum_{q=b+1}^{b+\tau_1-j} z^{N_0}_{q+j} (z^{N_0}_{q})^T 
		\end{bmatrix}.
	\end{aligned}
\end{equation}

We then estimate $\hbar_j$ by
\begin{equation}
	\begin{aligned}
		\hat{\hbar}_j &= \frac{\text{tr}_{\text{block}}(\mathcal{Z}_j)} {N_0 (\tau_1-j)} =\frac{1}{N_0 (\tau_1 - j)} \sum_{i=1}^{N_0} \sum_{q=b+1}^{ b + \tau_1-j} z^i_{q+j} (z^i_{q})^T .
	\end{aligned}
	\label{eq:h_hat}
\end{equation}
The estimate is unbiased, i.e., $E(\hat{\hbar}_j) = \hbar_j$, and the covariance of $\hat{\hbar}_j $ satisfies
\begin{equation}
	 \tau_1 \to \infty  \Rightarrow  \text{cov}(\hat{\hbar}_j) \to 0, 
\end{equation}
 similar to the proof in \cite{mehra1970identification}. Therefore, the estimation of $\Upsilon_{L}$, denoted as $\hat{\Upsilon}_{L}$, is
\begin{equation}
	\begin{aligned}
		& \hat{\Upsilon}_{L} = \begin{bmatrix}
			\hat{\hbar}_0^T & \hat{\hbar}_1^T & \dots & \hat{\hbar}_{L-1}^T \\
		\end{bmatrix}^T.
	\end{aligned}
\end{equation}
The estimation of $d$, denoted as $\hat{d} = \text{vec}( \hat{\Upsilon}_{L} )$, satisfies $ \hat{d} - d \sim \mathcal{O} (1/ \tau_1)$. Thus, the ALS algorithm, utilizing the estimated system model from \eqref{eq:est_model}, is defined as
\begin{equation}
	\begin{aligned}
	&	\hat{g} = \arg \min_{\tilde{g}} || \hat{\mathcal{A}} \tilde{g} - \hat{d} ||^2_2\\
		\text{s.t.} \quad &\hat{\mathcal{A}} = \begin{bmatrix}
			\hat{\mathcal{A}} _{1} & 	\hat{\mathcal{A}} _{2}
		\end{bmatrix},\\
	&\hat{\mathcal{A}} _{1} = (\hat{C} \otimes \mathcal{G}_o )(I_{n^2} - \bar{A} \otimes \bar{A})^{-1}, \\
			& \hat{\mathcal{A}} _{2} = 	\left[ \hat{\mathcal{A}} _{1}(\hat{A}K_f \otimes \hat{A}K_f) + I_{p} \otimes \mathcal{H} \right],\\
				& \tilde{g} = \begin{bmatrix}
				\text{vec}(\tilde{Q})^T & \text{vec}(\tilde{R})^T
			\end{bmatrix}^T. \\
	\end{aligned}
	\label{eq:optimaztion_propblem_of_ALS}
\end{equation}
This constitutes a standard least-squares problem, with a unique solution if and only if the matrix $\hat{\mathcal{A}}$ has full column rank.
Moreover, the true noise covariance matrices typically exhibit symmetry, with $\tilde{Q} \succeq 0$ and $\tilde{R} \succ 0$. Following the approach in \cite{rajamani2009estimation}, we apply the ``vech" operator to stack the column-wise elements from the lower triangular parts of $\tilde{Q}$ and $\tilde{R}$. This leads to the existence of two constant matrices, $D_q$ and $D_r$, such that $\text{vec}(\tilde{Q}) = D_q \text{vech}(\tilde{Q})$ and $\text{vec}(\tilde{R}) = D_r \text{vech}(\tilde{R} )$. Consequently, the optimization problem in \eqref{eq:optimaztion_propblem_of_ALS} is modified to 
\begin{equation}
	\begin{aligned}
		&	\hat{g}_{d} = \arg \min_{g} || \hat{\mathcal{A}_{d} } g_{d} - \hat{d} ||_2^2 \\
			\text{s.t.}  \quad &	\tilde{Q} \succeq 0, \tilde{R} \succ0,\\
		&\hat{\mathcal{A}}_d = \hat{\mathcal{A}} \begin{bmatrix}
		D_q^T & 	D_r^T
		\end{bmatrix}^T,\\
		&g_d = \begin{bmatrix}
			\text{vech}(\tilde{Q})^T & \text{vech}(\tilde{R})^T
		\end{bmatrix}^T.
	\end{aligned}
	\label{eq:optimaztion_propblem_of_ALS_new}
\end{equation}

\noindent \textbf{Remark 3.} The optimization problem in \eqref{eq:optimaztion_propblem_of_ALS_new} is a typical semidefinite programming problem (SDP), which can be solved using CVX, a package for formulating and solving convex programs \cite{cvx}. The estimated noise covariance matrices, $\hat{Q}$ and $\hat{R}$, are obtained from the optimal solution $\hat{g}_d$. The complete ADKF algorithm is presented in Algorithm~\ref{alg:alg1}.
\begin{algorithm}
	\caption{ADKF.}\label{alg:alg1}
	\begin{algorithmic}
		\STATE 
		\STATE {\textbf{Input: }}$Y,U,\hat{X}_0, L,\tau_1, Q_u,R_u$.
		\STATE {\textbf{Output: }}$ \hat{x}_{k|k} $.
		\STATE {\textbf{1) Compute the system matrices} $\hat{A},\hat{B}$, and $\hat{C}$}:
		\STATE \hspace{0.5cm}1. $\hat{A} = {G}_1^{\dagger} {G}_2 , \ \hat{B} = 	{G}_1^{\dagger} {G}_3$, and $\hat{{C}} = G_1(1:p;1:n)$,
		\STATE \hspace{0.5cm} $\ \ ${with} ${G}_1 , {G}_2$, and ${G}_3$ defined in \eqref{eq:g1g2g3}.
		
		\STATE {\textbf{2) Compute the noise covariance matrices} $\hat{Q}$ and $\hat{R}$}:
		\STATE \hspace{0.5cm}2. Compute the innovation sequence matrix $Z_s$ in \eqref{eq:newZ}.
		\STATE \hspace{0.5cm}3. $\hat{\hbar}_j = \frac{1}{N(\tau_1 - j)} \sum_{i=1}^{N} \sum_{q=b+1}^{b+\tau_1-j} z^i_{q+j} (z^i_{q})^T $.
		\STATE \hspace{0.5cm}4.	$ \hat{d}= \begin{bmatrix}
			\hat{\hbar}_0^T & \hat{\hbar}_1^T & \hat{\hbar}_2^T & \dots & \hat{\hbar}_{L-1}^T \\
		\end{bmatrix}^T$.\\
		\STATE \hspace{0.5cm}5. Compute $\hat{\mathcal{A}}_d $ defined in \eqref{eq:optimaztion_propblem_of_ALS_new}.
		\STATE \hspace{0.5cm}6. Solve \eqref{eq:optimaztion_propblem_of_ALS_new} to obtain $\hat{Q}$ and $\hat{R}$.
		\STATE {\textbf{3) State estimation}}:
		\STATE \hspace{0.5cm}7. \textbf{ For} $k = 0,1,\dots $ \textbf{do:} 
		\STATE \hspace{0.5cm}8. $\ \hat{x}_{k+1|k} = \hat{A} \hat{x}_{k|k} + \hat{B} u_{k} $.
		\STATE \hspace{0.5cm}9. $ \ P^a_{k+1|k} = \hat{A} P^a_{k|k} \hat{A}^T + \hat{Q} $.
		\STATE \hspace{0.5cm}10. $ K^a_k = P^a_{k|k-1} \hat{C}^T (\hat{C} P^a_{k|k-1} \hat{C}^T + \hat{R})^{-1} $.
		\STATE \hspace{0.5cm}11. $ \hat{x}_{k|k} = \hat{x}_{k|k-1} + K^a_k (y_k - \hat{C} \hat{x}_{k|k-1})$.
		\STATE \hspace{0.5cm}12. $P^a_{k|k} = P^a_{k|k-1} - K^a_k \hat{C} P^a_{k|k-1}$.
		\STATE \hspace{0.5cm}13. \textbf{End for} 
	\end{algorithmic}
	\label{alg1}
\end{algorithm}

\section{Performance Analysis of ADKF \label{sec5}}
In this section, we first explore the properties of the ALS method using the estimated system matrices. Next, we analyze the characteristics of the prior estimation error covariance matrix within the ADKF framework. Finally, we assess the posterior estimation performance of the proposed ADKF algorithm. 

\subsection{Performance of ALS with the Estimated System Model}
We begin by introducing a necessary lemma that will be used in the subsequent analysis.
 \vspace{6pt} 
 
\noindent \textbf{Lemma 1.}\cite[Theorem 1]{duan2023data}
Consider system \eqref{eq:LTImodel} with pre-collected data matrices $Y,U$, and $\hat{X}_0$ as defined in \eqref{eq:data_matrix}. Under Assumptions 1-3, for any positive scalar $\epsilon < \epsilon_0 $ and $\delta \in (0,1)$, there exists a constant $N_1$ such that if $N \ge N_1$, the following inequalities hold with probability at least $1 - \delta$
\begin{equation}
	\begin{aligned}
		||\Delta A ||_2 & = ||A - \hat{{A}}||_2 \le \epsilon, \\
		|| \Delta B||_2 & = ||B - \hat{{B}}||_2 \le \epsilon, \\
		|| \Delta C||_2 & = ||C - \hat{{C}}||_2 \le \epsilon, 
	\end{aligned}
\end{equation}
where $\epsilon \sim \mathcal{O} \left( \sqrt{\log(1/\delta)/ {N)}} \right)$, and 
\begin{equation}
	\epsilon_0  = \sqrt{||G_1||^2_2 + \lambda_{\min} (G^T_1 G_1)} - ||G_1||_2.
\end{equation}

Lemma 1 shows that a sufficiently large number of trajectories ensures that the estimation error of the system model converges to a small bound with a certain probability. From this, we can easily derive a corollary that establishes the relationship between $\tilde{Q}, \tilde{R}$ and $Q, R$.
 \vspace{6pt} 
 
\noindent \textbf{Corollary 1.}
Under the same condition specified in Lemma~1, for any positive scalar $\epsilon < \epsilon_0 $ and $ \delta \in (0,1)$, there exists a constant $N_1$ such that if $N \ge N_1$, the following inequalities hold with probability at least $1-\delta$
	\begin{equation}
		\begin{aligned}
		||\Delta \tilde{Q}||_2 & \triangleq ||\tilde{Q} - Q ||_2 \le \mathcal{O} \left(  \frac{ \log(1/\delta)}{N} \right), \\
		||\Delta \tilde{R}||_2 & \triangleq  ||\tilde{R} - R ||_2 \le \mathcal{O} \left( \frac{ \log(1/\delta)}{N} \right).  \\
		\end{aligned}
	\end{equation}
\begin{proof}
	See the proof in Appendix \ref{prof_corollary1}.\\
\end{proof}

For simplicity in the subsequent analysis of the ALS method with the estimated system model, we will ignore the constraints on the noise covariance matrices in \eqref{eq:optimaztion_propblem_of_ALS_new}. 
Based on this, we can now state the following results.
\vspace{6pt}

\noindent \textbf{Theorem 1.} Consider system \eqref{eq:LTImodel} with pre-collected data $Y,U,\hat{X}_0$, and $Z$, as defined in \eqref{eq:data_matrix} and \eqref{eq:collect_z}, respectively. Under Assumptions~1-4, if $\hat{\mathcal{A}}_d$ has full column rank, then the solution $\hat{g}_d$ of \eqref{eq:optimaztion_propblem_of_ALS} is unbiased, and the following inequalities hold
\begin{equation}
	\begin{aligned}
		|| \Delta \tilde{Q}^*||_2 &=  || \hat{Q} - \tilde{Q} ||_2   \le \mathcal{O} \left( \frac{1}{\sqrt{\tau_1}} \right), \\
		||\Delta \tilde{R}^*||_2 &=  || \hat{R} - \tilde{R} ||_2   \le \mathcal{O} \left( \frac{1}{\sqrt{\tau_1}} \right),\\
	\end{aligned}
\end{equation}
where $\hat{Q}$ and $\hat{R}$ are the estimated noise covariance matrices recovered from $\hat{g}_d$.

\begin{proof}
	See the proof in Appendix \ref{prof_prop2}.\\
\end{proof}
	
	\noindent \textbf{Remark 4.}
	Theorem~1 demonstrates that as $\tau_1 \to \infty$, the estimated $\hat{Q}$ and $\hat{R}$ converge to $\tilde{Q}$ and $\tilde{R}$, respectively. Consequently, $\hat{Q},\hat{R}$ are preferable to the true $Q$ and $R$ for filter design, as they account for inaccuracies in the system model.
	Additionally, Theorem~1 assumes that $\hat{\mathcal{A}_d}$ has full column rank, which ensures the unique existence of $\hat{g}_d$. We refer to this as ALS-estimatable. If $\hat{{A}}_d$ does not have full column rank, we can include additional terms in the cost function, such as $\rho ||g_d||_2$, to prevent overfitting. Assumption 4 is crucial for ensuring the unbiasedness of $\hat{g}_d$. Without this assumption, $\hat{g}_d$ will be biased, although this bias diminishes and converges to zero as $\tau_1 \to \infty$.
	\vspace{6pt}
	
	\noindent \textbf{Theorem 2.}
	Under the same condition specified in Theorem~1, for any positive scalar $\epsilon < \epsilon_0 $ and $ \delta \in (0,1)$, there exists a constant $N_1$, such that if $N \ge N_1$, the following inequalities hold with probability at least $1-\delta$
	\begin{equation}
		\begin{aligned}
			\Delta \hat{Q} & \triangleq ||\hat{Q} - Q||_2 \le  \mathcal{O} \left(  \frac{ \log(1/\delta)}{N} \right) + \mathcal{O} \left( \frac{1}{\sqrt{\tau_1}} \right),\\
			\Delta \hat{R} &\triangleq ||\hat{R} - R||_2 \le  \mathcal{O} \left(  \frac{ \log(1/\delta)}{N} \right) + \mathcal{O} \left( \frac{1}{\sqrt{\tau_1}} \right),
		\end{aligned}
	\end{equation}
	where $\hat{Q}$ and $\hat{R}$ are the estimated covariance matrices recovered from the solution $\hat{g}_d$ of \eqref{eq:optimaztion_propblem_of_ALS}.
	
	\begin{proof}
		See the proof in Appendix \ref{QQQQQQ}\\
	\end{proof}
	
			\noindent \textbf{Remark 5.} Theorem~2 establishes a sample-complexity bound for the convergence of the estimated covariance matrices $\hat{Q}$ and $\hat{R}$ to their true counterparts $Q$ and $R$. Specifically, we have $||\hat{Q} - Q|| \to 0$ and $||\hat{R} - R|| \to 0$ as both $N \to \infty$ and $\tau_1 \to \infty$.
\subsection{Properties of Priori Estimation Error Covariance}
		
For simplicity in notation, we assume $\hat{B} = B$ and $\hat{{C}} = C$ for the following analysis, as the performance analysis with $\hat{B}$ and $\hat{C}$ is identical to that with $\hat{A}$. Once the estimated system model is obtained, we denote the steady priori estimation error covariance in ADKF as $P_a$, which satisfies
\begin{equation}
	\begin{aligned}
		&P_a= \hat{A} (P_a^{-1}+ {C}^T \hat{R}^{-1} {C} )^{-1} \hat{A}^T +\hat{Q}.
	\end{aligned}
	\label{eq:p_f}
\end{equation}

Similarly, the steady priori estimation error covariance in i-KF, denoted as $P$, satisfies
\begin{equation}
	P = A(P^{-1} + C^T R^{-1} C)^{-1} A^T + Q.
	\label{eq:p_ideal}
\end{equation}
The difference between $P_a$ and $P$ is quantified as follows.
	\vspace{6pt}
	
\noindent \textbf{Proposition 2.	\label{theorem:PF_P_ERROR}}
Under the same condition specified in Theorem~1, for any positive scalar $\epsilon < \epsilon_0 $ and $ \delta \in (0,1)$, there exists a constant $N_1$ such that if $N \ge N_1$, the following inequality holds with probability at least $1-\delta$
\begin{equation}
	\begin{aligned}
		&||P_a-P||_2 \\
		&\leq   \mathcal{O} \left( \sqrt{ \frac{ \log(1/\delta)}{N}} \right)  + \mathcal{O} \left(  \frac{ \log(1/\delta)}{N} \right) + \mathcal{O} \left( \frac{1}{ \sqrt{\tau_1} } \right) .
	\end{aligned}
	\label{eq:Pa_P}
\end{equation}

\begin{proof}
	See the proof in Appendix \ref{append_pf_p}.\\
\end{proof}

\noindent \textbf{Remark 6.} Proposition 2 establishes the performance gap between $P_a$ and $P$ in terms of the number of trajectories $N$ and the hyperparameter $\tau_1$, with a probability of at least $1-\delta$. It is evident that $	||P_a-P||_2 \to 0$ as $N \to \infty$ and $\tau_1\to \infty$. 
 \vspace{6pt} 
 
Similarly, to compare the gap between NDKF, IDKF (NDKF with true $\tilde{Q}$ and $\tilde{R}$), and ADKF, the steady priori estimation covariance $P_{\sharp}$ in IDKF, and that in NDKF with nominal $Q_u$ and $R_u$, denoted as $P_f$, satisfy the following equations
\begin{subequations}
	\begin{align}
		P_{\sharp} & = \hat{A}(P_{\sharp}^{-1} + C^T R^{-1} C)^{-1} \hat{A}^T + {Q} \label{eq:p_sharp},\\
		P_{f} &= \hat{A}(P_{f}^{-1} + C^T R_u ^{-1} C)^{-1} \hat{A}^T + Q _u.
	\end{align}
\end{subequations}
 
\noindent \textbf{Corollary 2.} 
Under the same condition specified in Theorem~1, for any positive scalar $\epsilon < \epsilon_0 $ and $ \delta \in (0,1)$, there exists a constant $N_1$ such that if $N \ge N_1$, the following inequalities hold with probability at least $1-\delta$
\begin{subequations}
	\begin{align}
		|| P_{a} - P_{\sharp} ||_2 &\le  \mathcal{O} \left( \frac{1}{\sqrt{\tau_1}} \right),	\label{eq:pa_p}\\
	    || P_{f} - P ||_2 &\leq l_1||\Delta {Q}||_2 + l_2 ||\Delta {R}||_2 + l_3\epsilon,
	\label{eq:pa_p_sharp}
	\end{align}
\end{subequations} 
where $l_1$, $l_2$ and $l_3$ are constants, $ \Delta Q = Q_u - Q $, and $\Delta R = R_u - R$.

\begin{proof}
	See the proof in Appendix \ref{cor_pf_p_sharp}.\\
\end{proof}

Corollary 2 highlights that the primary difference between ADKF and IDKF arises from the mismatch between the true $\tilde{Q}$, $\tilde{R}$, and their estimates $\hat{Q}$ and $\hat{R}$. Notably, $||P_a -P_{\sharp}||_2 \to 0$ as $\tau_1 \to \infty$.
Inequality \eqref{eq:pa_p_sharp} indicates that the prediction performance of NDKF is primarily influenced by mismatched noise covariance. Simply increasing $N$ cannot eliminate the influence caused from mismatched noise covariances. As noted in \cite{ge2016performance}, whether $P_f \succeq P$ or $P \succ P_f $ depends on the characteristics of $\Delta Q$ and $\Delta R$. 

\subsection{ Estimation Performance Evaluation of ADKF}

The true posteriori estimation error covariance $P^m_{k|k}$ is different from the $P^a_{k|k}$ obtained in Algorithm~1. In this subsection, we evaluate the gap between $P^m_{k|k}$ and the ideal posteriori estimation error covariance obtained under the i-KF, denoted as $P_{k|k}$. Define the posterior state estimation error as $ \tilde{e}_{k|k} = \hat{x}_{k|k} - x_k $ and its corresponding error covariance is $ P^m_{k|k} \triangleq E\{\tilde{e}_k \tilde{e}_k^T \}$. The steady versions of $P^m_{k|k}$, $P^a_{k|k}$, and $P_{k|k}$ are denoted as $P^m_{\infty}$, $P^a_{\infty}$, and $P_{\infty}$, respectively. \\

\noindent \textbf{Theorem 3.}
Consider system \eqref{eq:LTImodel} with pre-collected data $Y,U,\hat{X}_0$, and $Z$, as defined in \eqref{eq:data_matrix} and \eqref{eq:collect_z}, respectively. Assume that $\hat{\mathcal{A}}_d$, defined in \eqref{eq:optimaztion_propblem_of_ALS}, has full column rank. Under Assumptions~1-4, for any positive scalar $\epsilon < \epsilon_0 $ and $ \delta \in (0,1)$, there exists a constant $N_1$ such that if $N \ge N_1$, the following inequality holds with probability at least $1 - \delta$
	\begin{equation}
		\begin{aligned}
			& \quad || P^m_{\infty} - P_{\infty}|| \\
			&\leq   \mathcal{O} \left( \sqrt{ \frac{ \log(1/\delta)}{N}} \right)  + \mathcal{O} \left(  \frac{ \log(1/\delta)}{N} \right) + \mathcal{O} \left( \frac{1}{\sqrt{\tau_1}} \right) .
		\end{aligned}
\end{equation}

\begin{proof}
See the proof in Appendix \ref{proof:proposend}.\\
\end{proof}

From the above analysis, we observe that the estimation performance of ADKF is influenced by the accuracy of the system model. As $N \to \infty$ and $\tau_1 \to \infty$, $|| P^m_{\infty} - P_{\infty}|| \to 0$, indicating that the proposed ADKF can achieve performance nearly equivalent to that of i-KF based on noisy data. This is based on having a sufficiently large number of trajectories and a sufficiently large $\tau_1$. Furthermore, since all estimation processes are derived from the data \eqref{eq:data_matrix} and \eqref{eq:collect_z}, the proposed ADKF algorithm is entirely data-driven.
\begin{figure*}
	\centering
	\begin{subfigure}[b]{0.32\textwidth} 
		\centerline{\includegraphics[width=\columnwidth]{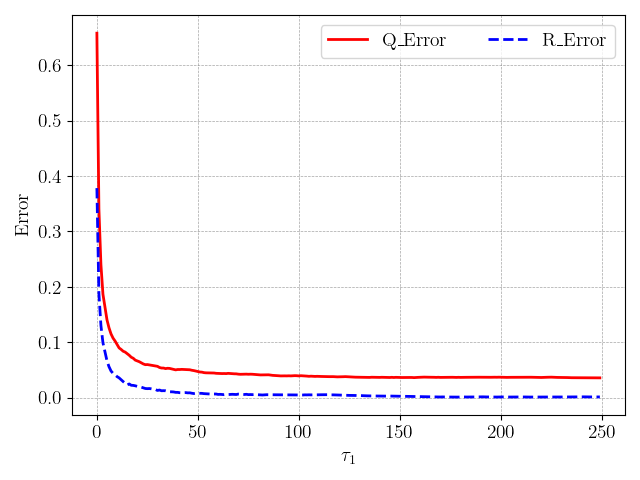}}
		\label{example1_L}
		\caption{}
	\end{subfigure}%
	\hfill 
	\begin{subfigure}[b]{0.32\textwidth} 
		\centerline{\includegraphics[width=\columnwidth]{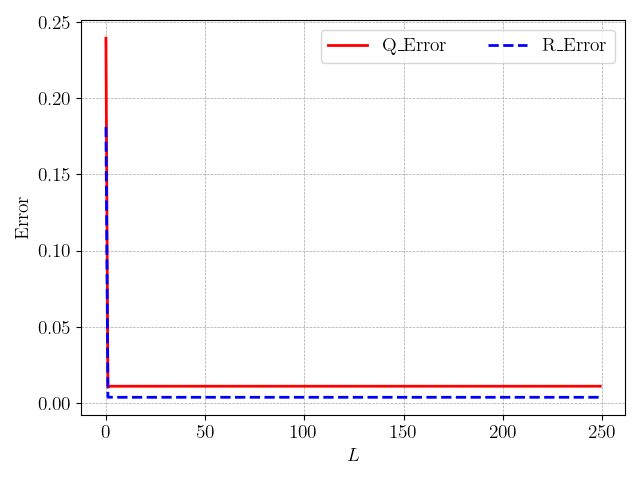}}
		\caption{}
		
	\end{subfigure}%
	\hfill 
	\begin{subfigure}[b]{0.32\textwidth} 
		\centerline{\includegraphics[width=\columnwidth]{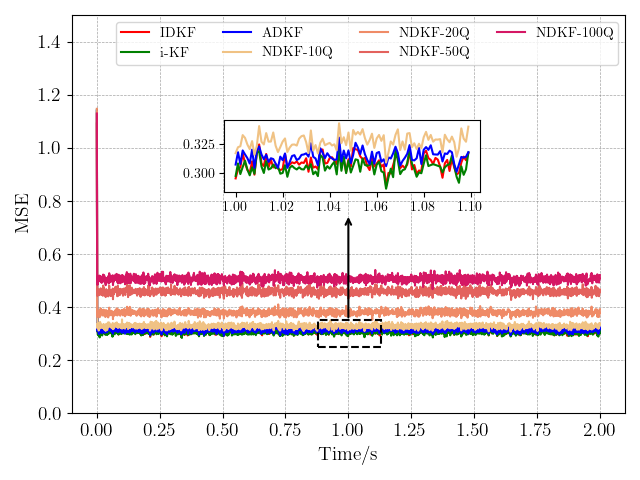}}
		\label{fig222}
		\caption{}
	\end{subfigure}
	\caption{DC Motor: (a) The estimation error of noise covariance matrices as $\tau_1$ increases, with $Q_u = 10Q, R_u = 5R$, $N = 5000$, $L=15$. (b) The estimation error of noise covariance matrices as $L$ increases, with $\tau_1 = 150$,$Q_u = 10Q, R_u = 5R$ and $N = 5000$. (c) MSE with $Q_u = \gamma Q$, $R_u = 5R$, $\gamma = [10.0, 20.0, 50.0, 100.0], L=15, \tau_1=200$.}
	\label{fig_L}
\end{figure*}

\section{Simulations \label{sec6}}
In this section, we present two examples to demonstrate the effectiveness of the proposed ADKF algorithm, and compare its filtering performance with three other algorithms: i-KF, IDKF, and NDKF. With a slight abuse of notation, we use IDKF to represents the filter that utilizes $\hat{A},\hat{B},\hat{C}$, and the true $Q,R$. To assess the performance, we define two types of estimation errors
\begin{equation}
	\begin{aligned}
		& \text{MSE}(N) = \frac{1}{N} \sum_{i=1}^{N} (x^i_k - \hat{x}^i_k)^2, \\
		& \text{AMSE}(k)= \frac{1}{100N} \sum_{i=1}^{N} \sum_{j=k}^{k+99} (x^i_j - \hat{x}^i_j)^2,
	\end{aligned}
\end{equation}
where $N$ denotes the number of Monte Carlo trials, $i$ refers to the $i$-th trial, and $k$ is the starting point. The MSE$(N)$ evaluates the mean squared error at a specific time $k$ across $N$ Monte Carlo trials. AMSE$(k)$ calculates the average mean squared error over the next 100 time steps starting from $k$, providing insight into the filter's performance over a period.

\subsection{DC Motor}

\begin{figure*}
	\centering
	\begin{subfigure}[b]{0.32\textwidth}
		\includegraphics[width=\textwidth]{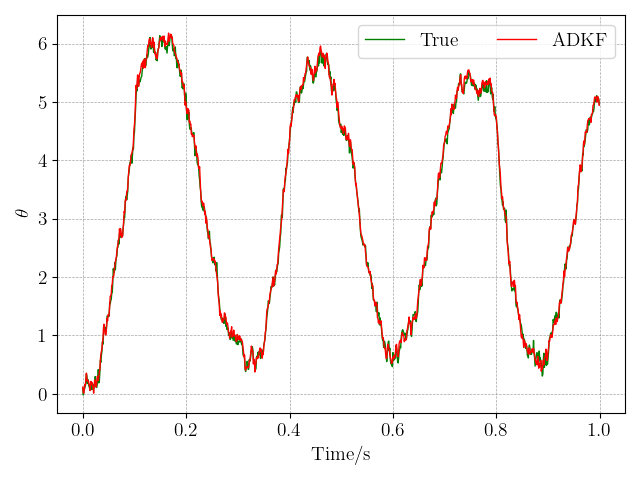}
		\caption{$\theta$}
		\label{fig:subfig1}
	\end{subfigure}
	\hfill
	\begin{subfigure}[b]{0.32\textwidth}
		\includegraphics[width=\textwidth]{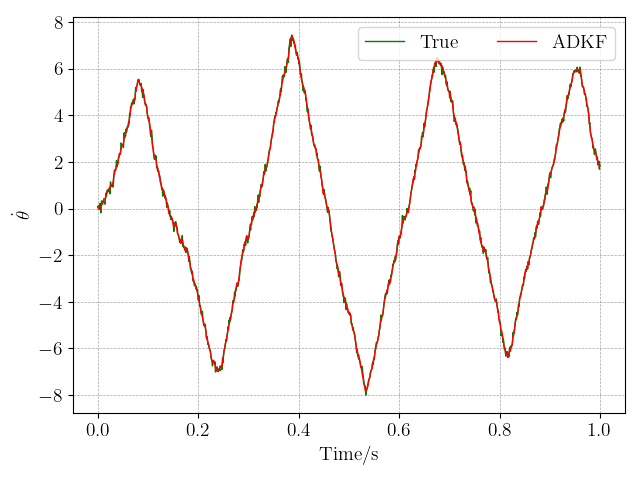}
		\caption{$\dot{\theta}$}
		\label{fig:subfig2}
	\end{subfigure}
	\hfill
	\begin{subfigure}[b]{0.32\textwidth}
		\includegraphics[width=\textwidth]{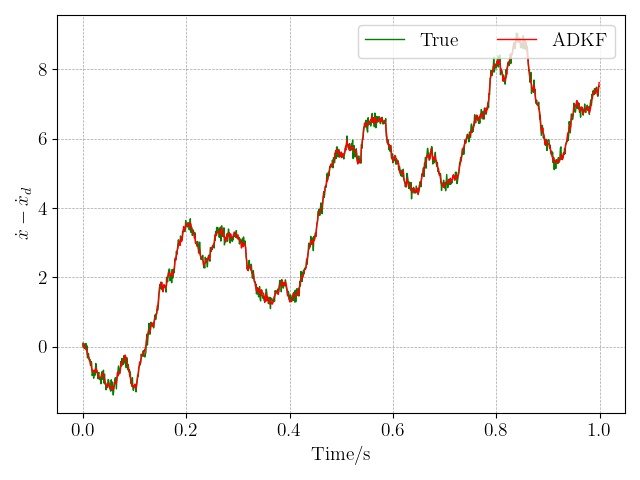}
		\caption{$\dot{x} - \dot{x}_d$}
		\label{fig:subfig3}
	\end{subfigure}
	\caption{Wheel-legged Robot: (a) $\theta$ is the tilt angle. (b) $\dot{\theta}$ is the angular velocity. (c) The gap between velocity and determinate velocity.}
	\label{fig_combined}
\end{figure*}

We utilize the DC motor system presented in \cite{shi2014event}, with a discrete-time system model described by
\begin{equation}
	{A} = {\begin{bmatrix}
			0.9951 & 0.2289\\
			-0.0177 & 0.8672
	\end{bmatrix}} , {B} ={\begin{bmatrix}
			-0.4158 & 0.0038\\
			-0.0038 & 0.0301
	\end{bmatrix}} ,
\end{equation}
and a sampling period $dt = 0.001s$. Let $x \triangleq
\begin{bmatrix} \dot{\theta} & i \end{bmatrix}^T$ and $ u \triangleq
\begin{bmatrix} T & V \end{bmatrix}^T$, where $ \dot{\theta}, i, T$, and $V$ represent the rotational velocity, current, load torque, and DC voltage input, respectively. The measurement matrix is $C =I_2$, and the true noise covariance matrices are give by 
\begin{equation}
	Q = \begin{bmatrix} 0.20 & 0.04 \\ 0.04 & 0.40 \end{bmatrix}, R = \begin{bmatrix} 0.50 & 0.01 \\ 0.01 & 0.50 \end{bmatrix}.
\end{equation}

Since this is a stable system, we only conduct 5000 independent experiments for data collection, each with $\tau = 1000$ time steps. The initial states and input are generated by setting $\sigma_x = \sigma_u = 100$, and $\Pi_0 = 0.1 I_n$. For noise covariance estimation, let $L = 20$ and $\tau_1 = 100$. The nominal noise covariance matrices are set as $Q_u = 5 Q$ and $R_u = \gamma R$, with $\gamma =[10, 20,50, 100]$ to compare the estimation performance of the proposed ADKF method under different nominal noise covariance matrices. 
   \vspace{6pt} 
   
As shown in Fig.~\ref{fig_L} (a) and (b), the estimation errors of noise covariance matrices remain relatively stable as $L$ increases but decrease as $\tau_1$ increases, aligning with the analysis mentioned earlier. In Fig.~\ref{fig_L} (c) and TABLE \ref{tab:tab1}, we observe that as the value of $\Delta Q$ increases, the MSE in NDKF also rises. However, the nominal setting of $Q_u$ has minimal impact on the estimation accuracy of the proposed ADKF method, which performs nearly identical to IDKF and i-KF. This indicates that ADKF effectively approximates the system model and the noise covariance matrices. Additionally, in TABLE \ref{tab:tab12}, we observe that as the nominal $R_u$ increases, the filtering performance of the NDKF method degrades sharply. In contrast, the ADKF method continues to improve estimation accuracy, though it remains slightly less accurate than IDKF and i-KF. This further demonstrates ADKF's robustness to initial parameter settings.
\begin{table}[H]
	\caption{AMSE under $R_u = 5R$ and different $Q_u$ in DC Motor.}
	\centering
	\begin{tabular}{cccccc}
		\toprule
		$Q_u$ & $10Q$ & $20Q$ & $50Q$ & $100Q$ \\
		\midrule
		{NDKF} & 0.326 & 0.379 & 0.460 & 0.509 \\
		\textbf{ADKF} & \textbf{0.311} & \textbf{0.310}& \textbf{0.309} & \textbf{0.309}\\
		{IDKF} & {0.308} & {0.308 }& {0.308} & {0.308}\\
		{i-KF} & {0.304} & {0.303 }& {0.305} & {0.303}\\
		\bottomrule
	\end{tabular}
	\label{tab:tab1}
\end{table}
\begin{table}[H]
	\caption{AMSE under $Q_u = 5Q$ and different $R_u$ in DC Motor.}
	\centering
	\begin{tabular}{cccccc}
		\toprule
		$R_u$ & $10R$ & $20R$ & $50R$ & $100R$ \\
		\midrule
		{NDKF} & 0.331 & 0.417 & 0.626 & 0.858 \\
		\textbf{ADKF} & \textbf{0.336} & \textbf{0.360 }& \textbf{0.401} & \textbf{0.432}\\
		{IDKF} & {0.304} & {0.304 }& {0.304} & {0.304}\\
		{i-KF} & {0.304} & {0.304 }& {0.304} & {0.304}\\
		\bottomrule
	\end{tabular}
	\label{tab:tab12}
\end{table}

\subsection{Wheel-Legged Robot}
In this subsection, we utilize a nonlinear system, the wheel-legged robot \cite{cui2021learning}, to illustrate the effectiveness of ADKF in nonlinear robotic system. 
\begin{figure}
	\centerline{\includegraphics[width=0.4 \columnwidth]{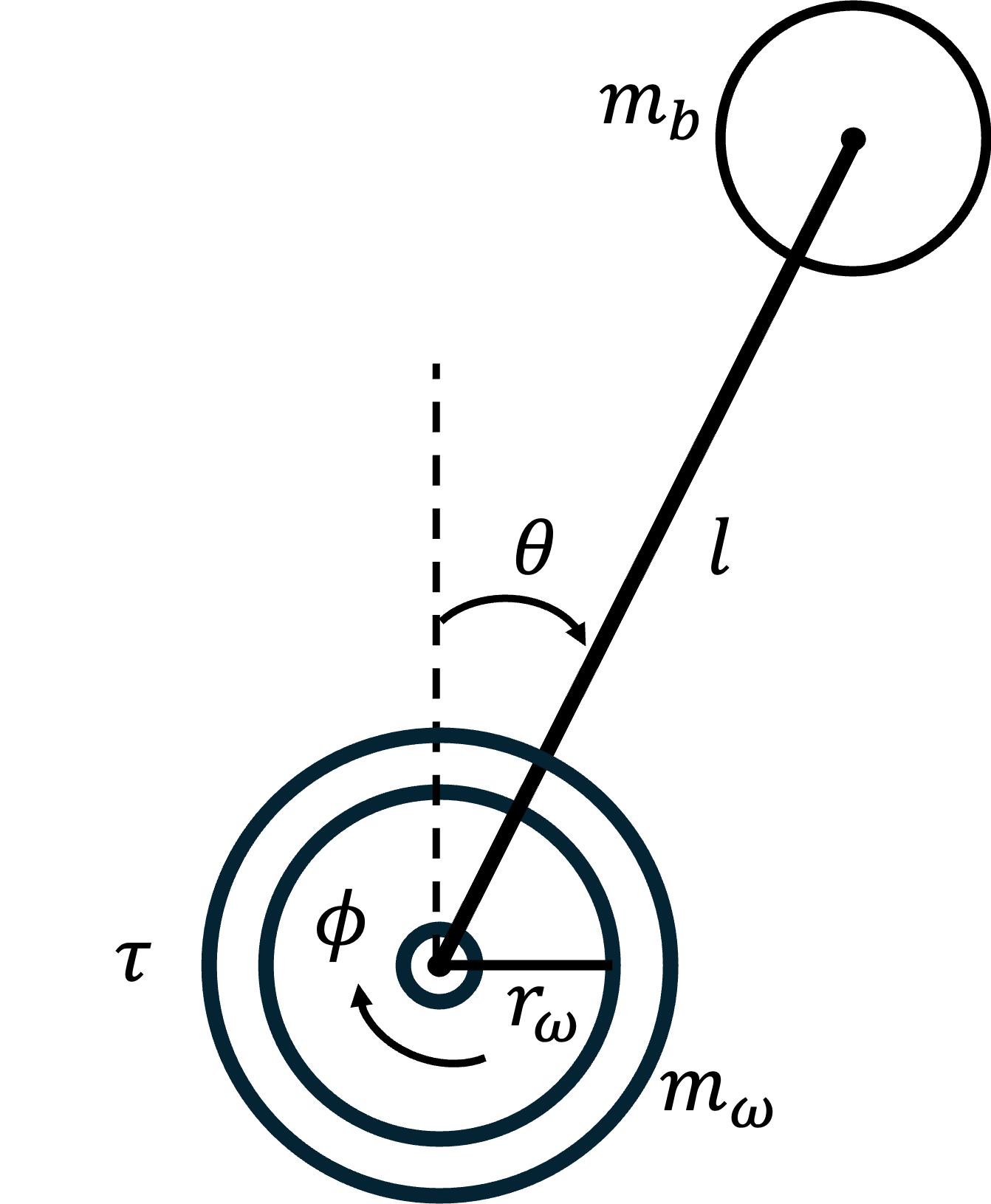}}
	\caption{Template model of the wheel-legged robot.}
	\label{fig:wheel_legged}
\end{figure}
As shown in Fig.~\ref{fig:wheel_legged}, the model of the wheel-legged robot can be simplified as an inverted pendulum, according to \cite{huang2010sliding}, the dynamic model is
\begin{equation}
	\begin{aligned}
		m_{11} \ddot{\phi} + m_{12} \cos(\theta) \ddot{\theta} &= m_{12} \sin(\theta) \dot{\theta}^2 + \tau, \\
		m_{12} \cos(\theta) \ddot{\phi} + m_{22} \ddot{\theta} &= G \sin(\theta) - \tau,
	\end{aligned}
	\label{eq:wheel_legged}
\end{equation}
where $m_{11} = (m_b + m_{\omega}) r^2_{\omega} + I_{\omega}, m_{12} = m_blr_{\omega}, m_{22} = m_b l^2 + I_b$, and $G =m_bgl$. Here, $m_{\omega}$ is the mass of the wheel, $m_b$ is the mass of the body, $I_{\omega}$ is the moment of inertia of the wheel, $I_b$ is the moment of inertia of the body, $\phi$ denote the rotational angle, $\theta$ is the tilt angle of the robot, $\tau$ denotes the torque of the wheel, $r_{\omega}$ is the radius of the wheel, and $l$ is the height of the body. Define $\dot{x} = \dot{\phi} r_{\omega}$ as the linear velocity of the robot. Then, the state of the wheel-legged robot can be defined as $ \boldsymbol{x} = [\theta, \dot{\theta}, \dot{x} - \dot{x}_d ]^T$, where $\dot{x}$ is the desired velocity. The dynamic model \eqref{eq:wheel_legged} can be written in the form of system~\eqref{eq:LTImodel}. Then, the matrix $C = I_3$ and the true noise covariance matrices are given by
\begin{equation}
	Q = \begin{bmatrix} 0.005 & 0.0 & 0.0 \\ 0.0 & 0.005 & 0.0\\0.0 & 0.0 & 0.005 \end{bmatrix}, \ R = \begin{bmatrix} 0.01 & 0.0 & 0.0 \\ 0.0 & 0.01 & 0.0\\0.0 & 0.0 & 0.01 \end{bmatrix}.
\end{equation}

We collect data based on the nonlinear model \eqref{eq:wheel_legged}, and discretize the system \eqref{eq:wheel_legged} using the fourth order Runge-Kutta method with a sampling period $\delta t = 0.01s$. We generate 5000 trajectories, with $\tau = 2000$, $\sigma_x = \sigma_u = 10$, and $\Pi_0 =0.01  I_n $. When estimating noise covariance, we generate another 100 trajectories and set $L =20, \tau_1=1000$, and $K_c = [1951, 386,	3]$. The initial estimation for noise covariance is set as $Q_u= 5 Q$, $R_u = \gamma R$, where $\gamma \in [1.0, 10.0, 50.0, 100.0]$. We compare the MSE among the ADKF, NDKF, IDKF, and Kalman filter with discrete local linear dynamic model and true noise covariances (i-KF). Furthermore, since the wheel-legged robot is a nonlinear system, we also compute the MSE in EKF method with true noise covariance matrices.
\begin{table}[H]
	\caption{AMSE under $Q_u=5Q$ and different $R_u$ in Wheel-Legged Robot.}
	\centering
	\begin{tabular}{cccccc}
		\toprule
		$R_u$ & $R$ &$10R$ & $50R$ & $100R$ \\
		\midrule
		{NDKF} & 0.111 & 2.494 & 13.902 & 26.797 \\
		\textbf{ADKF} & \textbf{0.017} & \textbf{0.018}& \textbf{0.018} & \textbf{0.019} \\
		{IDKF} & {1.038} & {1.038 }& {1.039} & {1.038} \\
		{EKF} & {0.015} & {0.015 }& {0.015} & {0.015} \\
		{i-KF} & {0.010} & {0.010 }& {0.010} & {0.010} \\
		\bottomrule
	\end{tabular}
	\label{tab2}
\end{table}

\begin{figure}
	\centerline{\includegraphics[width=0.95 \columnwidth]{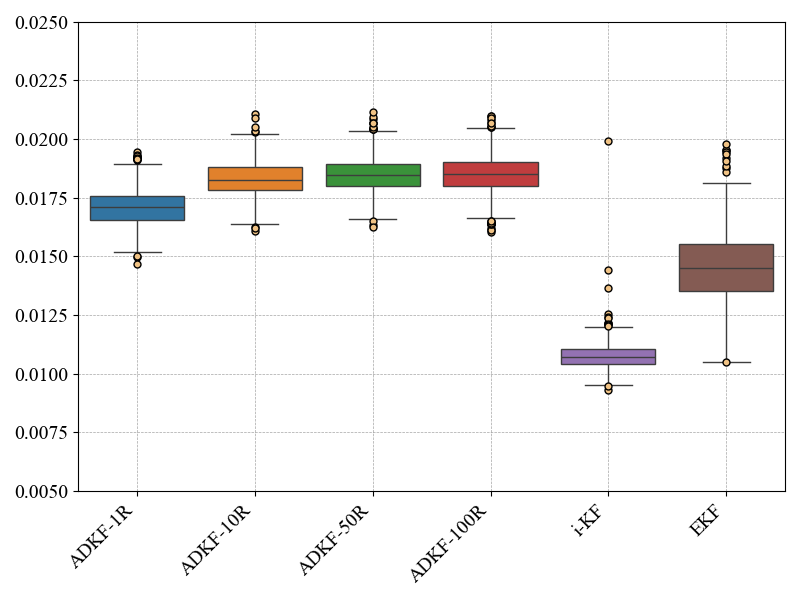}}
	\caption{MSE for three filters, with $R_u = \gamma R$, $Q_u = 5Q$, where $\gamma = [1.0, 10.0, 50.0, 100.0 ]$. }
\label{fig:subfig4}
\end{figure}

As illustrated in Fig.~\ref{fig:subfig4}, the proposed ADKF method outperforms both the NDKF and IDKF, which display values exceeding 0.1. TABLE \ref{tab2} shows that the initial settings of $Q_u$ and $R_u$ have minimal impact on the estimation performance of ADKF, further demonstrating its robustness. These results imply that the true noise covariance matrices are not well-suited for the filter design with estimated system model. We believe that ADKF effectively capture the properties the nonlinear system near the equilibrium point, providing more information and ultimately leading to better performance than other local-linearization-based methods. Furthermore, ADKF nearly matches the performance of the EKF and i-KF, despite a slightly higher MSE. However, unlike EKF and i-KF, which rely on accurate model, the ADKF estimates the state using only noisy data. Additionally, as a linear estimator, ADKF guarantees stability and is significantly more computationally efficient than EKF.

\section{Conclusion \label{sec7}}
This article has addressed the data-driven state estimation problem for unknown systems. The proposed ADKF algorithm simultaneously identifies both the linear system model and the noise covariance matrices from noisy pre-collected data, significantly improving estimation performance compared to the NDKF method with mismatched noise covariance matrices. The ADKF algorithm can achieve nearly the same performance as the optimal Kalman filter with an accurate system model. Although the filter performance demonstrated in this article is limited to LTI systems, simulations suggest that the ADKF method has the potential to handle nonlinear systems, as the estimated noise covariance matrices account for system model inaccuracies, making the algorithm more suitable for estimation problems in unknown systems. In future work, we intend to apply the proposed algorithm to robotic systems, combining the part known system information into the design, which is also an important topic.

\section{Appendix}
%
%
\subsection{Proof of Proposition 1 \label{pppp1}}
	Under the control gain $K_c$, the system model in \eqref{eq:LTImodel} can be rewritten as 
\begin{equation}
	\begin{aligned}
		x_k 
		& = (A+BK_cC) ^k x_0  + \sum_{i=1}^{k} (A+BK_cC)^{i-1} \omega_{k-i}.
	\end{aligned}
\end{equation}
From the initialization of $x_0$ as described in Section \ref{data_collection}, we we can express the expected value of $x_kx_k^T$ as
\begin{equation}
	\begin{aligned}
		E \{x_k x_k^T\} 
		& = (A+BK_cC) ^k (\Pi_0 + \sigma_x I_n) [ (A+BK_c C) ^k]^T  \\
		& \quad + \sum_{i=1}^{k} (A+BK_cC)^{i-1} Q [(A + BK_cC)^{i-1}]^T. \\
	\end{aligned}
\end{equation}

Since $A + BK_cC$ is Schur stable, as $k \to \infty$, we have $(A + BK_cC)^k \to 0$. Therefore, according to \cite[Lemma~2.1]{anderson2005optimal} as $k \to \infty $, we obtain
\begin{equation}
	\begin{aligned}
		E\{x_kx_k^T\}  
		&=  \sum_{i=1}^{\infty} (A+BK_c C)^{i-1} Q [(A + BK_c C)^{i-1 }]^T \\
		&\triangleq \Pi_1,
	\end{aligned}
	\label{eq:e_xk_xk}
\end{equation}
where $\Pi_1$ is a constant matrix. Utilizing \eqref{eq:e_xk_xk}, we can express $\Delta \tilde{Q}_k$ for $k\to \infty$ as
\begin{equation}
	\begin{aligned}
		\Delta \tilde{Q}_k &= (\Delta A + \Delta B K_c C) E\{x_k x_k^T\}  (\Delta A + \Delta B K_c C)^T\\
		&  =(\Delta A + \Delta B K_c C) \Pi_1 (\Delta A + \Delta B K_c C)^T \triangleq \Pi_2,
	\end{aligned}
	\label{eq:pi_2}
\end{equation}
where $\Pi_2$ is also a constant matrix.  Additionally, according to \cite{qian2023observation} (Lemma 3, Theorem 2), there exists a constant $b_0$ such that $\sum_{i=0}^{\infty} ||(A+BK_c C)^{i}||_2 ||(A+BK_c C)^i||_2 \le b_0$. Therefore, we have
\begin{equation}
	\begin{aligned}
		|| \Delta \tilde{Q}_k||_2 \le b_1 || Q||_2,
	\end{aligned}
	\label{eq:l1}
\end{equation}
where $b_1 = b_0||\Delta A + \Delta B K_c C||^2_2 $.
	\vspace{6pt}
	
Similarly, we can express
\begin{equation}
	\begin{aligned}
		\Delta \tilde{R}_k = \Delta C \Pi_1 \Delta C^T \triangleq \Pi_3,
	\end{aligned}
	\label{eq:pi_3}
\end{equation}
where $\Pi_3$ is a constant matrix. We also can obtain
\begin{equation}
	|| \Delta \tilde{R}_k||_2 \le b_2 || Q||_2,
		\label{eq:l2}
\end{equation}
where $b_2 = b_0 || \Delta C||_2^2 $.

\subsection{Proof of Corollary 1 \label{prof_corollary1}}
The proof of Corollary 1 is similar regardless of whether $A$ is stable or unstable. Therefore, we focus on analyzing $\tilde{Q}$ obtained under the feedback controller $K_c$ when $A$ is not stable. From \eqref{eq:e_xk_xk}, it follows that $E\{ x_k x_k^T\} = \Pi_1 $ as $ k \to \infty$. Consequently, there exists a positive definite matrix $M$ such that 
\begin{equation}
	E\{ x_k x_k^T\} \preceq M, k = 0,\dots,\tau.
\end{equation}

Based on \eqref{eq:QTK} and Lemma 1, for $\epsilon < \epsilon_0$ and $N \ge N_1$, the following inequality holds with probability $1-\delta$
\begin{equation}
		\begin{aligned}
		&\quad ||\tilde{Q} - Q||_2  \\
			& = (\Delta A + \Delta B K_c C) E\{x_k x_k^T\}  (\Delta A + \Delta B K_c C)^T\\
			&\le ||M||_2 || \Delta A + \Delta B K_c C ||_2^2 \\
			&\le a_1 \epsilon^2 \le  \mathcal{O} \left(  \frac{ \log(1/\delta)}{N} \right),
		\end{aligned}
\end{equation}
where 
\[a_1 =  || M ||_2 (1 + ||K_c C||_2^2 ).\]

Similarly, we obtain $||\tilde{R} - R||_2 \le  \mathcal{O} \left( { \log(1/\delta)} / {N} \right)$ with probability at least $1-\delta$, when $\epsilon < \epsilon_0$ and $N \ge N_1$.

	 \subsection{Proof of Theorem 1 \label{prof_prop2}}
Based on the unbiased properties of $\hat{d}$ as discussed in Section~\ref{als_fram}, the expectation of the estimated $\hat{g}_d$ is given by
 \begin{equation}
 	\begin{aligned}
 		E(\hat{g}_d) &= \hat{\mathcal{A}}_d^{\dagger } E(\hat{d})= \hat{\mathcal{A}}_d^{\dagger } d \\
 		& = \hat{\mathcal{A}}_d^{\dagger } \hat{\mathcal{A}}_d g = g.
 	\end{aligned}
 \end{equation}
 The covariance of the estimated $\hat{g}_d$ is
 \begin{equation}
 	\text{cov}(\hat{g}_d) = \hat{\mathcal{A}}_d^{\dagger } \text{cov}({\hat{d}} ) (\hat{\mathcal{A}}_d^{\dagger })^T.
 \end{equation}
 
 As $\text{cov}(\hat{d}) = E \{ (\hat{d} - d \}) (\hat{d} - d \})^T\} \sim \mathcal{O}(1/\tau_1)$, by the Cauchy–Schwarz inequality, we obtain that
 \begin{equation}
 	E\{ ||\hat{d} - d||_2   \} \le  \sqrt{ E\{ ||\hat{d} - d||^2 _2 \} } \le \mathcal{O} ( 1 / \sqrt{\tau_1}).
 \end{equation}
 Thus, we can further deduce
 \begin{equation}
 	\begin{aligned}
 		|| \hat{g}_d - g_d ||_2 &= ||   \hat{\mathcal{A}}_d^{\dagger } \hat{d} -  \hat{\mathcal{A}}_d^{\dagger } d ||_2 \\
 		&= ||   \hat{\mathcal{A}}_d^{\dagger } (\hat{d} - d) ||_2  \le \mathcal{O} ( 1 / \sqrt{\tau_1}).\\
 	\end{aligned}
 \end{equation}
 
 Additionally, we have
 \begin{equation}
 	\begin{aligned}
 		| |\hat{g}_d - g_d ||_2^2& =\begin{Vmatrix}
 			\text{vech}(\hat{Q} - \tilde{Q})  \\
 			\text{vech}( \hat{R} - \tilde{R} ) 
 		\end{Vmatrix}_2^2 \\
 		&= || \text{vech}(\hat{Q} - \tilde{Q}) ||_2^2 +  ||  \text{vech}(\hat{R} - \tilde{R} ) ||_2^2 \\
 		& = \frac{ || \hat{Q} - \tilde{Q}  ||^2_F + || \text{diag} (\hat{Q} - \tilde{Q}) ||^2_F  }{2} \\
 		&\quad + \frac{ ||\hat{R} - \tilde{R}  ||^2_F + || \text{diag} (  \hat{R} - \tilde{R} ) ||^2_F  }{2} \\
 		& \le \mathcal{O} \left ( \frac{1} { \tau_1} \right ).\\
 	\end{aligned}
 \end{equation}
 Therefore, the estimated $\hat{Q}$ and $\hat{R}$, recovered from $\hat{g}_d$, satisfy
 \begin{equation}
 	\begin{aligned}
 		\Delta \tilde{Q}^* &=  || \hat{Q} - \tilde{Q} ||_2 \le || \hat{Q} - \tilde{Q} ||_F   \le \mathcal{O} \left( \frac{1}{\sqrt{\tau_1}} \right), \\
 		\Delta \tilde{R}^* &=  || \hat{R} - \tilde{R} ||_2 \le || \hat{R} - \tilde{R} ||_F  \le \mathcal{O} \left( \frac{1}{\sqrt{\tau_1}} \right).\\
 	\end{aligned}
 	\label{eq:q_hat_q_tilde}
 \end{equation}

\subsection{Proof of Theorem 2 \label{QQQQQQ}}
		Based on Lemma~2, when $N \ge N_1$ and $\epsilon < \epsilon_0$, the following holds with probability at least $1-\delta$
\begin{equation}
			\begin{aligned}
	\Delta \hat{Q} & =  ||  \hat{Q} - Q ||_2 \le ||  \hat{Q} - \tilde{Q} ||_2 +  || \tilde{Q} - Q ||_2 \\
	&\le  \mathcal{O} \left(  \frac{ \log(1/\delta)}{N} \right) + \mathcal{O} \left( \frac{1}{\sqrt{\tau_1}} \right).
\end{aligned}
\end{equation}
Similarly, we have 
\begin{equation}
\begin{aligned}
	\Delta \hat{R}  \le  \mathcal{O} \left(  \frac{ \log(1/\delta)}{N} \right) + \mathcal{O} \left( \frac{1}{\sqrt{\tau_1}} \right),
\end{aligned}
\end{equation}
which holds with probability at least $1-\delta$, when $N \ge N_1$ and $\epsilon < \epsilon_0$.\\

\subsection{Proof of Proposition 2 \label{append_pf_p}}
Following from (\ref{eq:p_f}) and (\ref{eq:p_ideal}), we have
\begin{equation}
	\begin{aligned}
		P_a- P &= \hat{A} (P_a^{-1}+ C^T \hat{R}^{-1} C) ^{-1} \hat{A}^T \\
		& \quad  - A(P^{-1} + C^T R^{-1} C)^{-1} A^T + \hat{Q} - Q .
		\label{eq:P_a_P}
	\end{aligned}
\end{equation}
\\
Let the first two terms on the right side of the formula in (\ref{eq:P_a_P}) be denoted by $\Delta P_{\sharp}^a$, so we can write
\begin{equation}
	\begin{aligned}
		&\quad \ \Delta P_{\sharp}^a \\
		&= \hat{A} ( P_a^{-1}+ C^T\hat{R}^{-1} C )^{-1} \hat{A}^T - \hat{A} (P_a^{-1}+ C^T \hat{R}^{-1} C ) ^{-1} {A}^T \\
		& +\hat{A}(P^{-1} + C^T R^{-1} C)^{-1} {A}^T - {A}(P^{-1} + C^T R ^{-1} C)^{-1} A^T \\
		& + \hat{A}(P_a^{-1}+ C^T\hat{R}^{-1} C)^{-1} {A}^T -\hat{A}(P^{-1} + C^T R^{-1} C)^{-1} {A}^T.
	\end{aligned}
	\label{eq:Delta_P_sharP_a}
\end{equation}

The first four terms in (\ref{eq:Delta_P_sharP_a}), denoted as $\Delta P_{\sharp,1}^a$, satisfy the following inequality
\begin{equation}
	\begin{aligned}
		 \Delta P_{\sharp,1}^a&= \hat{A}(P_a^{-1}+ C^T \hat{R}^{-1} C)^{-1} (\hat{A}^T - A^T)\\
		&\quad \quad + (\hat{A} - A)(P^{-1} + C^T R^{-1} C)^{-1} {A}^T \\
		&\leq ||\hat{A}(P_a^{-1}+ C^T \hat{R}^{-1} C)^{-1}||_2 ||\hat{A} - A||_2 I_n\\
		&\quad  + ||(P^{-1} + C^T R^{-1} C)^{-1} {A}^T||_2 ||\hat{A} - A||_2I_n \\
		&\leq a_2 ||\hat{{A}} - A||_2 I_n,
	\end{aligned}
\end{equation}
where
\begin{equation}
	\begin{aligned}
		a_2&= ||\hat{A}(P_a^{-1}+ C^T \hat{R}^{-1} C)^{-1}||_2 \\
		&\quad + ||(P^{-1} + C^T R^{-1} C)^{-1} {A}^T||_2 .\\
	\end{aligned}
\end{equation}
 For the last two terms in (\ref{eq:Delta_P_sharP_a}), denoted as $\Delta P_{\sharp,2}^a$, we have
\begin{equation}
	\begin{aligned}
    & \quad \ \Delta P_{\sharp,2}^a\\
    &= \hat{A} \left[ (P_a^{-1} + C^T \hat{R}^{-1} C)^{-1} - (P^{-1} + C^T R^{-1} C)^{-1} \right ] {A}^T \\
		& = \hat{A}(P_a^{-1} + C^T \hat{R}^{-1} C)^{-1} \left[ P^{-1} - P_a^{-1} + C^T (R^{-1} \right. \\
		&\left. \quad - \hat{R}^{-1})C \right ] (P^{-1} + C^T R^{-1} C)^{-1} {A}^T \\
		& = \hat{A}(P_a^{-1} + C^T \hat{R}^{-1} C)^{-1} P_a^{-1}(P_a - P) P^{-1} \\
		&\quad \times(P^{-1} + C^T R^{-1} C)^{-1} {A}^T + \hat{A}(P_a^{-1} + C^T \hat{R}^{-1} C)^{-1} \\
		&\quad \times C^T \hat{R}^{-1} \Delta \hat{R} R^{-1} C (P^{-1} + C^T R^{-1} C)^{-1} {A}^T. \\
	\end{aligned}
\end{equation}

Introducing the following notations
\begin{equation}
	\begin{aligned}
		\hat{F}_{a} & \triangleq \hat{A}(P_a^{-1} + C^T \hat{R}^{-1} C)^{-1} P_a^{-1},\\
		F  &\triangleq A (P^{-1} + C^T R^{-1} C)^{-1} P^{-1}, \\
		H & \triangleq P_a C^T \hat{R}^{-1} \Delta \hat{R} R^{-1} C P.
	\end{aligned}
\end{equation}
We can express $\Delta P_{\sharp,2}^a$ as
\begin{equation}
	\begin{aligned}
		 \Delta P_{\sharp,2}^a = \hat{F}_{a} (P_a - P ) {F}^T + \hat{F}_{a} H {F}^T.
	\end{aligned}
\end{equation}
Therefore, equation \eqref{eq:P_a_P} can be rewritten as 
\begin{equation}
	\begin{aligned}
	 P_a - P &= \Delta P^a_{\sharp,1} + \hat{F}_{a} (P_a - P ) {F}^T + \hat{F}_{a} H {F}^T + \Delta \hat{Q}\\
	& = \Delta P^a_{\sharp,1} + \hat{F}_{a} \Delta P^a_{\sharp,1} {F}^T + \hat{F}_{a}^2 (P_a - P ) ({F}^T)^2\\
	&\quad + \hat{F}_{a}^2 H ({F}^T)^2 + \hat{F}_{a} H {F}^T +\hat{F}_{a} Q {F}^T + \Delta \hat{Q} \\
	& \quad \quad \quad \quad \vdots \\
	& = \sum_{i=0}^{\infty}\hat{F}_{a}^{i} \Delta P^a_{\sharp,1} ({F}^T)^i + \hat{F}_{a}^{\infty} (P_a - P ) ({F}^T)^{\infty} \\
	&\quad + \sum_{i=1}^{\infty}\hat{F}_{a}^{i} H ({F}^T)^i 
	 + \sum_{i=0}^{\infty}\hat{F}_{a}^{i} \Delta \hat{Q} ({F}^T)^i .
	\end{aligned}
	\label{eq:Pf_P_error_eq1}
\end{equation}
Furthermore, we have 
\begin{equation}
	\begin{aligned}
			 \hat{F}_a &=\hat{A}(P_a^{-1} + C^T R^{-1} C)^{-1} P_a^{-1}\\
			 & = 	\hat{A}\left[ (P_a - P_a C^T(R+ C P_a C^T)^{-1} C P_a \right] P_a^{-1}\\
			 &= \hat{A} (I_n - K_aC).
	\end{aligned}
\end{equation}
which is Schur stable. Similarly, $F$ is also Schur stable, implying that the term $ \hat{F}_{a}^{\infty} (P_a - P ) ({F}^T)^{\infty} = 0$ in (\ref{eq:Pf_P_error_eq1}). Consequently, we have
\begin{equation}
	\begin{aligned}
		&P_a - P = \sum_{i=0}^{\infty}\hat{F}_{a}^{i} \Delta P^a_{\sharp,1} ({F}^T)^i + \sum_{i=1}^{\infty}\hat{F}_{a}^{i} H ({F}^T)^i \\
		&\quad \quad \quad \quad 
		+ \sum_{i=0}^{\infty}\hat{F}_{a}^{i} \Delta \hat{Q} ({F}^T)^i .
	\end{aligned}
	\label{eq:P_a_P_error}
\end{equation}

According to \cite{qian2023observation} (Lemma 3, Theorem 2), there exists a constant $a_3$ such that $\sum_{i=0}^{\infty}||\hat{F}_{a}^{i}||_2 ||{F}^i||_2 \le a_3$.
Thus, the norm $||P_a -P||_2$ is upper bounded by
\begin{equation}
	\begin{aligned}
		& ||P_a-P||_2 \leq a_3 (||H||_2 + ||\Delta \hat{Q}||_2 + a_2 || \hat{A} - A||_2 ).
	\end{aligned}
	\label{eq:PF_p_upper_bound}
\end{equation}

Since $||H||_2 \leq ||P_a C^T \hat{R}^{-1} ||_2 || R^{-1} C P||_2 ||\Delta \hat{R}||_2 $, it follows from Lemma 1 and Theorem~1, that for any positive scalar $\epsilon < \epsilon_0$ and $\delta \in (0,1)$, there exists a constant $N_1$ such that when $N \ge N_1$, the inequality
\begin{equation}
	\begin{aligned}
	& \quad ||P_a-P||_2 \\
	 &\leq a_3||\Delta \hat{Q}||_2 + a_4 ||\Delta\hat{R}||_2 +a_2a_3\epsilon \\
	& \leq   \mathcal{O} \left( \sqrt{ \frac{ \log(1/\delta)}{N}} \right)  + \mathcal{O} \left(  \frac{ \log(1/\delta)}{N} \right) + \mathcal{O} \left( \frac{1}{ \sqrt{\tau_1}} \right),
	\end{aligned}
	\label{eq:||pf_p||}
\end{equation}
 holds with probability at least $1-\delta$, where $a_4= a_3|| P_a C^T \hat{R}^{-1} ||_2 || R^{-1} C P ||_2 $.

\subsection{Proof of Corollary 3 \label{cor_pf_p_sharp}}
Following from \eqref{eq:p_f} and \eqref{eq:p_sharp}, the error $P_a - P_{\sharp}$ can be expressed as 
\begin{equation}
	\begin{aligned}
		&\quad P_{a} - P_{\sharp} \\
		& =\hat{A} \left[ (P_{a} ^{-1} + C^T \hat{R}^{-1} C)^{-1} - (P_{\sharp}^{-1} + C^T \tilde{R}^{-1} C)^{-1}\right] \hat{A}^T \\
		&\quad + \Delta \tilde{Q}^* \\
		&= \hat{A}(P_{a} ^{-1} + C^T \hat{R}^{-1} C)^{-1} 
		\left[ P_{\sharp}^{-1} - P_{a} ^{-1} + C^T ( \tilde{R}^{-1} \right. \\
		& \left. \quad - \hat{R}^{-1})C \right] (P_{\sharp}^{-1} + C^T \tilde{R}^{-1} C)^{-1} \hat{A}^T + \Delta \tilde{Q}^*  \\
		& = \hat{A}(P_{a} ^{-1} + C^T \hat{R}^{-1} C)^{-1} P_{a} ^{-1}(P_{a} - P_{\sharp}) P_{\sharp}^{-1} (P_{\sharp}^{-1} \\
		& \quad + C^T \tilde{R}^{-1} C)^{-1} \hat{A}^T + \hat{A}(P_{a} ^{-1} + C^T \hat{R}^{-1} C)^{-1} C^T \hat{R}^{-1} \\
		& \quad \times \Delta \hat{R} \tilde{R}^{-1} C (P_{\sharp}^{-1} + C^T \tilde{R}^{-1} C)^{-1} \hat {A}^T + \Delta \tilde{Q}^* .
	\end{aligned}
	\label{eq:pa_p_s}
\end{equation}

Introducing the following notations
\begin{equation}
	\begin{aligned}
		&F_{\sharp} \triangleq \hat{A} (P_{\sharp}^{-1} + C^T \tilde{R}^{-1} C)^{-1} P_{\sharp}^{-1},\\
		&H_{\sharp} \triangleq P_{a} C^T R_u^{-1} \Delta \hat{R} \tilde{R}^{-1} C P_{\sharp},
	\end{aligned}
\end{equation}
we can rewritten \eqref{eq:pa_p_s} as
\begin{equation}
	\begin{aligned}
		P_{a} - P_{\sharp} &= \hat{F}_{a}(P_{a} - P_{\sharp}) {F}_{\sharp}^T + \hat{F}_{a} H_{\sharp} \hat{F}_{\sharp}^T +\Delta \tilde{Q}^* \\
		& \quad \quad \vdots \\
		& = \hat{F}_{a}^{\infty} (P_{a} - P_{\sharp} ) ({F}_{\sharp}^T)^{\infty} + \sum_{i=1}^{\infty}\hat{F}_{a}^{i} {H}_{\sharp} ({F}_{\sharp}^T)^i \\
		& \quad + \sum_{i=0}^{\infty}\hat{F}_{a}^{i} \Delta \tilde{Q}^* ({F}_{\sharp}^T)^i . \\
	\end{aligned}
\end{equation}
Since $F_{\sharp}$ is also Schur stable, we conclude
\begin{equation}
			P_{a} - P_{\sharp}= \sum_{i=1}^{\infty}\hat{F}_{a}^{i} {H}_{\sharp}  ({F}_{\sharp}^T)^i + \sum_{i=0}^{\infty}\hat{F}_{a}^{i} \Delta \tilde{Q}^*  ({F}_{\sharp}^T)^i.\\
			\label{eq:P_a_p_sharp_error}
\end{equation}
Then, the upper bound for (\ref{eq:P_a_p_sharp_error}) is
\begin{equation}
	\begin{aligned}
		|| P_{a} - P_{\sharp} ||_2 
		&\le a_5 || \Delta \tilde{Q}^* ||_2 + a_6 ||{\Delta \tilde{R}^* } ||_2 .
		\label{eq:a4a5}
	\end{aligned}
\end{equation}
where $a_5$ is a constant such that $ \sum_{i=0}^{\infty}||\hat{F}_{a}^{i} ||_2 || ({F}_{\sharp}^T)^i ||_2 \le a_5$, and $a_6 = a_5 || P_{a} C^T \hat{R}^{-1}||_2 || \tilde{R}^{-1} C P_{\sharp}||_2$. Therefore, following from \eqref{eq:q_hat_q_tilde}, for any positive scalar $\epsilon < \epsilon_0$ and $\delta \in (0,1)$, there exists a constant $N_1$ such that when $N \ge N_1$, the inequality
\begin{equation}
	\begin{aligned}
		|| P_{a} - P_{\sharp} ||_2 \le  \mathcal{O} \left( \frac{1}{\sqrt{\tau_1}} \right),
	\end{aligned}
\end{equation}
holds with probability at least $1-\delta$.
 \vspace{6pt} 
 
The proof of inequality \eqref{eq:pa_p_sharp} follows a similar approach to that in Appendix \ref{append_pf_p}. Consequently, we can derive a result similar to \eqref{eq:P_a_P_error}: for any positive scalar $\epsilon < \epsilon_0$ and $\delta \in (0,1)$, there exists a constant $N_1$ such that when $N \ge N_1$, the following inequality holds with probability at least $1-\delta$
\begin{equation}
	 || P_{f} - P ||_2 \leq a_7||\Delta {Q}||_2 + a_8 ||\Delta {R}||_2 + a_9\epsilon,
\end{equation}
where $a_7,a_8$ and $a_9$ are constants.

\subsection{Proof of Theorem 3 \label{proof:proposend}}

The state estimation error is defined as $\tilde{e}_{k|k} = \hat{x}_{k|k} - x_k $. Then,
\begin{equation}
	\begin{aligned}
		\tilde{e}_{k+1|k+1} &= \hat{x}_{k+1|k+1} - x_{k+1} \\
		& = (I_n - K_{a} C)\hat{A}e_{k|k} + (I_n - K_{a} C)(\hat{A} - A)x_k \\
		& \quad  - (I_n - K_{a} C) {\omega}_k + K_{a} {\nu}_k.
	\end{aligned} 
\end{equation}
Since $P^m_{k|k} = E\{\tilde{e}_{k|k} \tilde{e}_{k|k}^T\}$, according to \cite{duan2023data} and Assumption~5, we have 
\begin{equation}
	\begin{aligned}
		P^m_{k+1|k+1} &\leq (1 + \eta)(I_n - K_{a} C) \hat{A} P^m_{k|k} \hat{A} ^T (I_n - K_{a} C)^T \\
		&\quad + K_{a} R K_{a}^T  + (I_n - K_{a} C)Q(I_n - K_{a} C)^T \\
		&\quad + (1 + 1/\eta)(I_n - K_{a} C)(\hat{A} - A) M \\
		&\quad \times (\hat{A}- A)^T (I_n - K_{a} C)^T,
	\end{aligned}
\end{equation}
where $\eta$ is a positive scalar and 
\begin{equation}
	\eta < \min \left\{ \|\hat{A} - A\|_2, \frac{1}{|\lambda((I_n - K_a C) \hat{A})|^2 - 1} \right\}.
\end{equation}

Next, we analyze the error between $P^m_{k+1|k+1}$ and $P^a_{\infty}$. For a Joseph form covariance update, the $P^a_{\infty}$ satisfies
\begin{equation}
	\begin{aligned}
		P^a_{\infty} & = (I_n - K_aC) P_{a} (I_n - K_aC)^T + K_a \hat{R} K_a^T\\ 
		&= (I_n - K_aC)( \hat{{A}} P^a_{\infty} \hat{{A}}^T + \hat{Q})(I_n - K_aC)^T + K_a \hat{R} K_a^T.
	\end{aligned}
\end{equation}
Moreover, from Assumption 5, we have $E\{ x_kx_k^T\} \le M $. Therefore, we have  
	\begin{equation}
		\begin{aligned}
			& \quad P^m_{k+1|k+1}- P^a_{\infty} \\
			& \le 
			(I_n - K_{a} C) \hat{A} (P^m_{k|k}-P^a_{\infty} ) \hat{A} ^T (I_n - K_{a} C)^T \\
			& \quad+ (1 + \frac{1}{\eta})(I_n - K_{a} C)(\hat{A} - A) M (\hat{A}- A)^T (I_n - K_{a} C)^T \\
			& \quad+ \eta (I_n - K_{a} C) \hat{A}P^m_{k|k} \hat{A} ^T (I_n - K_{a} C)^T + K_a \Delta \hat{R} K_a^T \\
			&\quad + (I_n -   K_{a} C)\Delta \hat{Q} (I_n - K_{a} C)^T\\
			& \le (I_n - K_{a} C) \hat{A} (P^m_{k|k}-P^a_{\infty} ) \hat{A} ^T (I_n - K_{a} C)^T \\
			&\quad + a_{10} ||\hat{A} - A||_2 I_n + a_{11} || \Delta \hat{Q} ||_2 I_n + a_{12} || \Delta \hat{R} ||_2 I_n \\
			& \quad \quad \quad \quad \vdots\\
			& \le [(I_n - K_{a} C) \hat{A} ]^{k+1} (P^m_{0|0} -P^{+}_{\sharp} )[ \hat{A} ^T (I_n - K_{a} C)^T ]^{k+1} \\
			&\quad + (a_{10} ||\hat{A} - A|| _2 + a_{11} || \Delta \hat{Q} ||_2  + a_{12} || \Delta \hat{R} ||_2)\\
			&\quad \times \sum_{i=0}^{k} [(I_n - K_{a} C) \hat{A} ]^i [ \hat{A} ^T (I_n - K_{a} C)^T]^i,
		\end{aligned}
	\end{equation}
	where 
\[
\begin{aligned}
			a_{10} & = (1 + 1/\eta)||(I_n - K_{a} C)||_2 || M ||_2 || (I_n - K_{a} C)^T ||_2 \\
&\quad \times  ||\Delta A|| _2 + || (I_n - K_{a} C) \hat{A} P_{e,k} \hat{A} ^T (I_n - K_{a} C)^T||_2, \\
a_{11} & = ||I_n-K_aC||_2^2, \text{ and } a_{12} = ||K_a||^2_2.
\end{aligned}
\]

Similarly, there exists a constant $a_{13}$, such that $\sum_{i=0}^{\infty} [(I_n - K_{a} C) \hat{A} ]^i [ \hat{A} ^T (I_n - K_{a} C)^T]^i \le a_{13} I_n$. As $k \to \infty $, we can obtain that
\begin{equation}
	P_{e, \infty} - P^a_{\infty} \le a_{14} || \Delta \hat{Q} ||_2 + a_{15} || \Delta \hat{R} ||_2 + a_{16} \epsilon,
\end{equation}
where $a_{14} = a_{11}a_{13}$, $a_{15} = a_{12} a_{13}$, and $a_{16} = a_{10}a_{13}$.

Additionally, 
\begin{equation}
	\begin{aligned}
		& \quad P^a_{\infty} - P_{\infty} \\
		& = (P_a + C^T \hat{R} ^{-1} C)^{-1} - (P + C^T {R} ^{-1} C)^{-1} \\
		& = (P_a + C^T \hat{R} ^{-1} C)^{-1} ( P - P_a) (P + C^T {R} ^{-1} C)^{-1} \\
		& \quad + (P_a + C^T \hat{R} ^{-1} C)^{-1} C^TR^{-1} \Delta\hat{R} \hat{R}^{-1} C\\
		&\quad \times (P + C^T {R} ^{-1} C)^{-1}.
	\end{aligned}
\end{equation}
Therefore, according to Theorem~1 and Proposition~1, we have
\begin{equation}
	\begin{aligned}
		||P^a_{\infty} - P_{\infty} ||_2  \le a_{17} ||\Delta \hat{Q}||_2 + a_{18}||\Delta \hat{R}||_2+ a_{19} \epsilon,
	\end{aligned}
\end{equation}
where
\[
\begin{aligned}
 a_{20} & =  ||(P_a + C^T \hat{R} ^{-1} C)^{-1}||_2 ||(P + C^T {R} ^{-1} C)^{-1}||_2, \\
 a_{21} & =  a_{20} ||C^TR^{-1}||_2 ||\hat{R}^{-1}C||_2 , \ a_{17} = a_3 a_{20}, \\
 a_{18} &=  a_4 a_{20} + a_{21}, \ a_{19} = a_5 a_{20}.
\end{aligned}
\]
Then we have
\begin{equation}
	\begin{aligned}
		||P_{e, \infty} - P_{\infty}||_2 &= ||P_{e, \infty} - P^a_{\infty}+ P^a_{\infty}- P_{\infty}||_2\\
		& \le ||P_{e, \infty} -P^a_{\infty} ||_2 + 	|| P^a_{\infty} - P_{\infty}||_2\\
		&\le a_{22} ||\Delta \hat{R}||_2 + a_{23}||\Delta \hat{Q}||_2+ a_{24} \epsilon,
	\end{aligned}
\end{equation}
where $a_{22}, a_{23}$, and $a_{34}$ are constants, and
\[
\begin{aligned}
	a_{22} &= a_{14} + a_{17},\
	a_{23} = a_{15}+a_{18}, \
	a_{24} = a_{16}+a_{19}.
\end{aligned}
\]

Therefore, according to Lemma~1 and Theorem~1, we have 
\begin{equation}
	\begin{aligned}
		& \quad ||P_{e, \infty} - P_{\infty}||_2 \\
		& \leq   \mathcal{O} \left( \sqrt{ \frac{ \log(1/\delta)}{N}} \right)  + \mathcal{O} \left(  \frac{ \log(1/\delta)}{N} \right) + \mathcal{O} \left( \frac{1}{\sqrt{\tau_1}} \right),
	\end{aligned}
\end{equation}
holds with probability at least $1-\delta$, when $N \ge N_1$ and $\epsilon < \epsilon_0$.


\section*{References}
\bibliographystyle{IEEEtran}
\bibliography{ref.bib}

\begin{IEEEbiography}[{\includegraphics[width=1in,height=1.25in,clip,keepaspectratio]{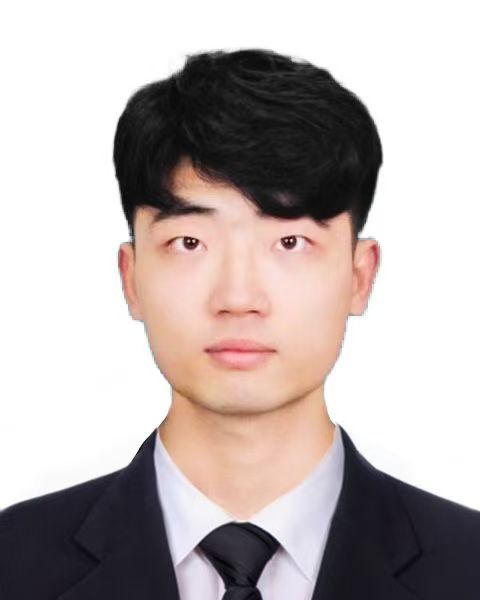}}]{Suyang Hu} 
	Suyang Hu received the B.Eng. and M.Eng. degree in Aerospace Science and Technology from Northwestern Polytechnical University, Xi'an, China, in 2021 and 2023, respectively. He is currently pursuing the Ph.D. degree in Electronic and Computer Engineering at the Hong Kong University of Science and Technology, Hong Kong, China. His current research interests include networked estimation and control, data-driven state estimation and control, and multi-robot systems.
\end{IEEEbiography}

\begin{IEEEbiography}[{\includegraphics[width=1in,height=1.25in,clip,keepaspectratio]{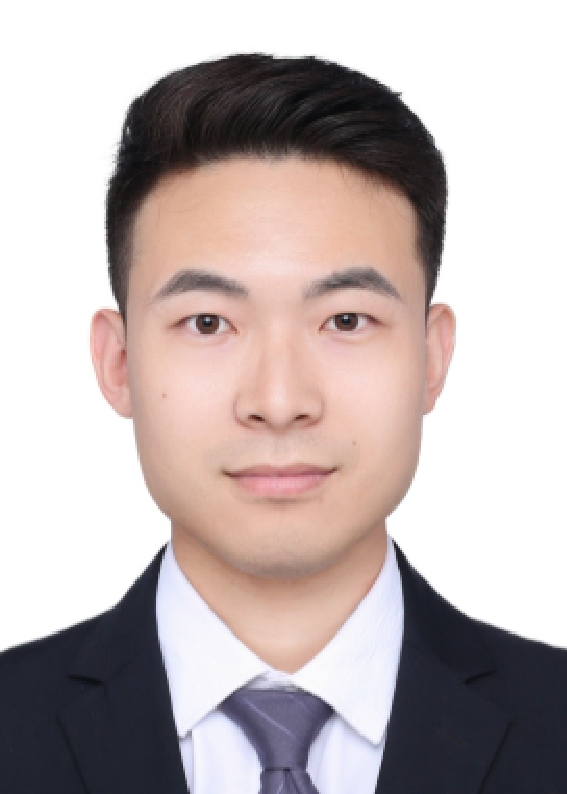}}]{Xiaoxu Lyu} 
Xiaoxu Lyu received the B.Eng. degree in Naval Architecture and Marine Engineering from Harbin Institute of Technology, Weihai, China, in 2018, and the Ph.D. degree in Dynamical Systems and Control from Peking University, Beijing, China, in 2023.

He is currently a Postdoctoral Fellow with the Department of Electronic and Computer Engineering, Hong Kong University of Science and Technology, Hong Kong, China. From September 2023 to December 2023, he was a Research Assistant with the Department of Systems Science, Southeast University, China. His research interests include data-driven state estimation and control, networked estimation and control, and multi-agent robotic systems.
\end{IEEEbiography}

\begin{IEEEbiography}[{\includegraphics[width=1in,height=1.25in,clip,keepaspectratio]{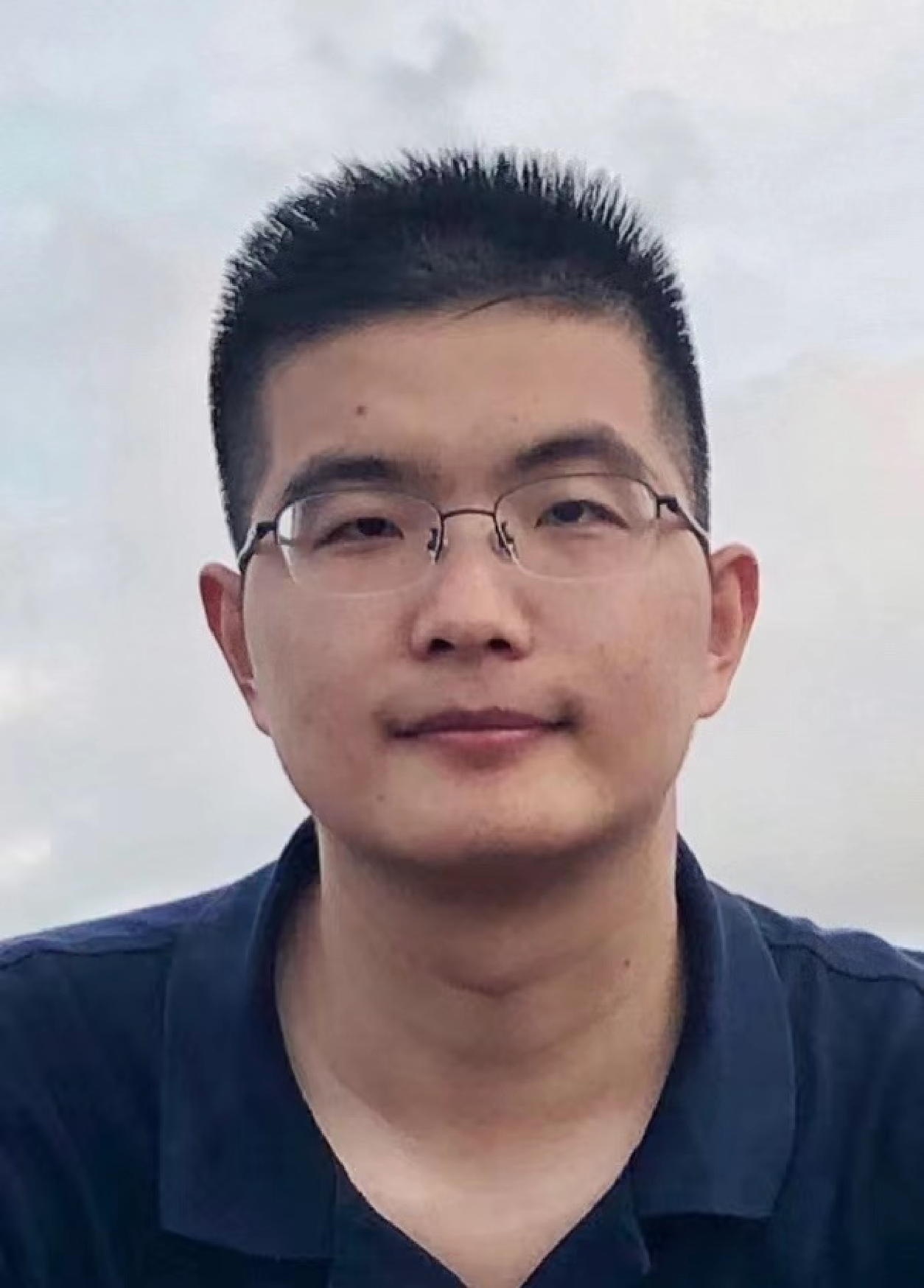}}]{Peihu Duan} 
Peihu Duan received the B.Eng. degree in Mechanical Engineering from Huazhong University of Science and Technology, Wuhan, China, in 2015. He received the Ph.D. degree in Mechanical Systems and Control from Peking University, Beijing, China, in 2020. Currently, he works as a Postdoc at School of Electrical Engineering and Computer Science, KTH Royal Institute of Technology, Stockholm, Sweden. From October 2020 to October 2022, He was a Postdoc at the Hong Kong University of Science and Technology and the University of Hong Kong, Hong Kong, China, respectively. His research interests include cooperative and data-driven control and state estimation of networked systems.
\end{IEEEbiography}

\begin{IEEEbiography}[{\includegraphics[width=1in,height=1.25in,clip,keepaspectratio]{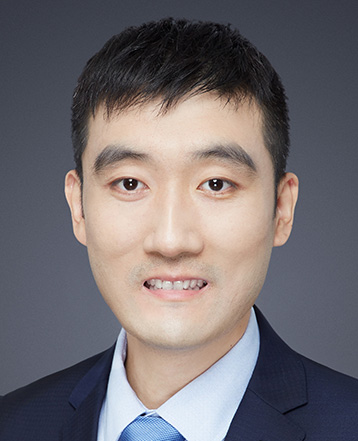}}]{Dawei Shi} 
(Senior Member, IEEE) Dawei Shi received the B.Eng. degree in electrical engineering and its automation from the Beijing Institute of Technology, Beijing, China, in 2008, and the Ph.D. degree in control systems from the University of Alberta, Edmonton, AB, Canada, in 2014. In December 2014, he was appointed as an Associate Professor with the School of Automation, Beijing Institute of Technology. From February 2017 to July 2018, he was with the Harvard John A. Paulson School of Engineering and Applied Sciences, Harvard University, Cambridge, MA, USA, as a Postdoctoral Fellow in bioengineering. Since July 2018, he has been with the School of Automation, Beijing Institute of Technology, where he is a Professor. His research focuses on the analysis and synthesis of complex sampled-data control systems with applications to biomedical engineering, robotics, and motion systems.

Dr. Shi is an Associate Editor/Technical Editor for IEEE Transactions on Industrial Electronics, IEEE/ASME Transactions on Mechatronics, IEEE Control Systems Letters, and IET Control Theory and Applications. He is a Member of the Early Career Advisory Board of Control Engineering Practice. He was the Guest Editor for European Journal of Control. He was an Associate Editor for the IFAC World Congress and is a Member of the IEEE Control Systems Society Conference Editorial Board.
\end{IEEEbiography}

\begin{IEEEbiography}[{\includegraphics[width=1in,height=1.25in,clip,keepaspectratio]{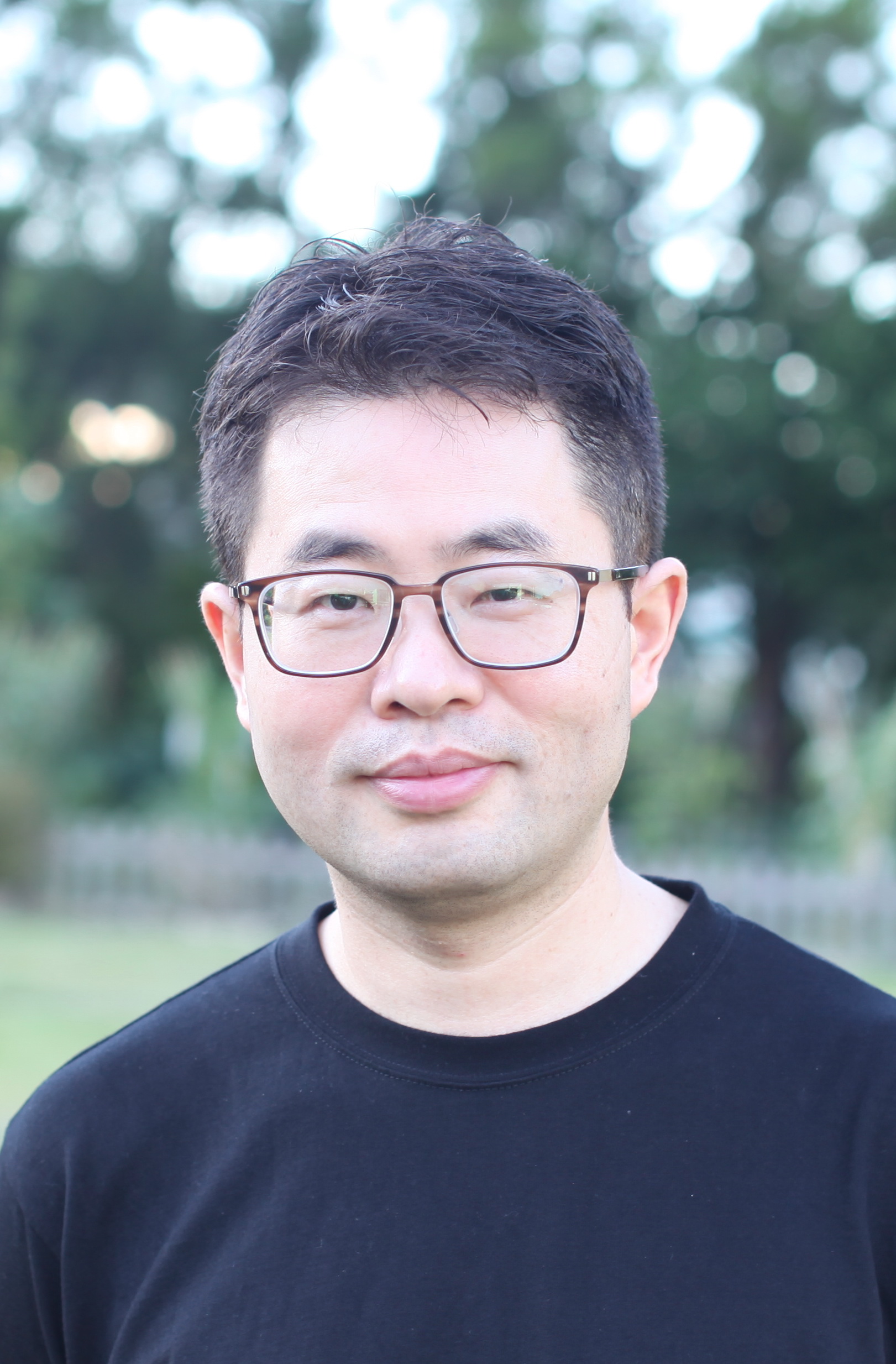}}]{Ling Shi} (Fellow, IEEE) Ling Shi received his B.E. degree in Electrical and Electronic Engineering from The Hong Kong University of Science and Technology (HKUST) in 2002 and  Ph.D. degree in Control and Dynamical Systems from The California Institute of Technology (Caltech) in 2008. He is currently a Professor in the Department of Electronic and Computer Engineering at HKUST. His research interests include cyber-physical systems security, networked control systems, sensor scheduling, event-based state estimation, and multi-agent robotic systems (UAVs and UGVs). He served as an editorial board member for the European Control Conference 2013-2016. He was a subject editor for International Journal of Robust and Nonlinear Control (2015-2017), an associate editor for IEEE Transactions on Control of Network Systems (2016-2020), an associate editor for IEEE Control Systems Letters (2017-2020), and an associate editor for a special issue on Secure Control of Cyber Physical Systems in IEEE Transactions on Control of Network Systems (2015-2017). He also served as the General Chair of the 23rd International Symposium on Mathematical Theory of Networks and Systems (MTNS 2018). He is currently serving as a member of the Engineering Panel (Joint Research Schemes) of the Hong Kong Research Grants Council (RGC). He received the 2024 Chen Fan-Fu Award given by the Technical Committee on Control Theory, Chinese Association of Automation (TCCT, CAA). He is a member of the Young Scientists Class 2020 of the World Economic Forum (WEF), a member of The Hong Kong Young Academy of Sciences (YASHK), and he is an IEEE Fellow.
\end{IEEEbiography}

\end{document}